\documentclass[pdflatex,sn-mathphys-num]{sn-jnl}

\usepackage{graphicx}\usepackage{multirow}\usepackage{amsmath,amssymb,amsfonts}\usepackage{amsthm}\usepackage{mathrsfs}\usepackage[title]{appendix}\usepackage{xcolor}\usepackage{textcomp}\usepackage{manyfoot}\usepackage{booktabs}\usepackage{algorithm}\usepackage{algorithmicx}\usepackage{algpseudocode}\usepackage{listings}\usepackage{bm}
\usepackage{tikz}
\usetikzlibrary{arrows.meta,calc}
\tikzset{
  pt/.style={circle,draw,inner sep=1.2pt,fill=white},
  arr/.style={-Latex,thick},
  lbl/.style={font=\small}
}

\theoremstyle{thmstyleone}\newtheorem{theorem}{Theorem}\newtheorem{proposition}[theorem]{Proposition}\newtheorem{lemma}[theorem]{Lemma}
\newtheorem{corollary}[theorem]{Corollary}

\theoremstyle{thmstyletwo}\newtheorem{remark}{Remark}

\theoremstyle{thmstylethree}\newtheorem{definition}{Definition}\newtheorem{assumption}[theorem]{Assumption}
\newtheorem{problem}[theorem]{Problem}
\begin{document}

\title[Long--Moody construction and multiplicative middle convolution]{Long--Moody construction of braid group representations and Haraoka's
multiplicative middle convolution for KZ-type equations}

\author*{\fnm{Haru} \sur{NEGAMI}}\email{negamiharu@gmail.com}

\affil{\orgdiv{Smart Sensing Technology Development Center},\\ \orgname{Research Institute
for Marine Technology and Engineering,\\ Japan Agency for Marine-Earth Science and Technology (JAMSTEC)
}, \orgaddress{\street{2-15 Natsushima-cho},
  \city{Yokosuka},
  \state{Kanagawa},
  \country{Japan}}}

\abstract{In this paper, we establish a correspondence between the algebraic and analytic approaches to constructing representations of the braid group $B_n$: the Katz-Long-Moody construction and the multiplicative middle convolution for Knizhnik-Zamolodchikov (KZ)-type equations, respectively. Furthermore, we demonstrate that this construction preserves the representations' unitarity relative to a Hermitian matrix and present an algorithm that determines the signature of this matrix for all but finitely many parameters $\lambda$ with $|\lambda|=1$ and $\lambda\neq 1$.
}



\maketitle

\section{Introduction}

Braid group representations play a central role in modern mathematics, with far-reaching applications in topology, representation theory, and mathematical physics. This profound significance arises from the fact that braid groups can be defined from multiple perspectives, notably through both algebraic and topological frameworks.

In 1925, Artin defined the braid group $B_n$ as the group of isotopy classes of structures formed by the intertwining of $n$ strings \cite{Artin1925}. 

\begin{definition}[The Artin braid group $B_n$]
The Artin braid group $B_n$ is the group generated by the $n-1$ generators $\sigma_1, \cdots , \sigma_{n-1}$ subject to the following two braid relations:

$$
\begin{array}{lcll}
    \text{1. }& \sigma_i\sigma_j &=\sigma_j\sigma_i\ \phantom{|} &(|j-i|>1) \\
    \text{2. } &\sigma_{i}\sigma_{i+1}\sigma_{i} 
    &=\sigma_{i+1}\sigma_{i}\sigma_{i+1}\ \ &(i=1, \cdots, n-2)
\end{array}
$$
\end{definition}

Birman provided a unified perspective on braid groups, mapping class groups, and knot theory \cite{Birman1974}. The braid group can also be defined as the mapping class group of an \( n \)-punctured closed disk. The generators of this group are half-twists, which are self-homeomorphisms that exchange two points.
\begin{definition}[Mapping class group]
    Let \(D_n\) denote a closed disk with \(n\) punctures, i.e., the closed disk \(D\) with \(n\) interior points, $p_1, \dots, p_n$, removed. The braid group is defined as the mapping class group of \(D_n\) that fixes the boundary \(\partial D_n\):
\[
B_n \cong \mathrm{MCG}(D_n) = \mathrm{Homeo}^{+}(D_n, \partial D_n) \big/ \sim,
\]
where $\mathrm{Homeo}^{+}(D_n, \partial D_n)$ denotes the group of orientation-preserving homeomorphisms $f \colon D_n \to D_n$ with $f|_{\partial D_n} = \mathrm{id}$, and \(\sim\) denotes the equivalence relation by isotopy.
\end{definition}
\begin{definition}[half-twist]
    The standard generators of $\mathrm{MCG}(D_n)$ are given by the \emph{half-twists}, which exchange adjacent punctures. More precisely, for each \(i=1,\dots,n-1\), define the half-twist \(h_i : D_n \to D_n\) as follows:
\begin{itemize}
  \item In a small neighborhood containing the adjacent punctures \(a_i\) and \(a_{i+1}\), the map \(h_i\) rotates the neighborhood by an angle of \(\pi\), thereby exchanging \(a_i\) and \(a_{i+1}\).
  \item This map is extended continuously to the entire \(D_n\) while fixing the boundary \(\partial D\).
\end{itemize}
The half-twists \(\{ h_1, h_2, \dots, h_{n-1} \}\) generate $\mathrm{MCG}(D_n)$ and satisfy the braid relations. 
\end{definition}

The pure braid group $P_n$ is a particularly important subgroup, defined as the kernel of the canonical projection from $B_n$ to the symmetric group $S_n$.

\begin{definition}[Pure braid group]
Let $S_n$ be the symmetric group on $n$ letters. The pure braid group \( P_n \) is defined as the kernel of the surjective homomorphism
\[
\Pi \colon B_n \to S_n, \quad \sigma_i \mapsto (i, i+1),
\]
which sends each standard Artin generator \( \sigma_i \) of the braid group \( B_n \) to the adjacent transposition \( (i, i+1) \in S_n \). 
A set of generators for \( P_n \) is given by
\[
\sigma_{ij} := \sigma_i \cdots \sigma_{j-2} (\sigma_{j-1})^2 (\sigma_i \cdots \sigma_{j-2})^{-1}, \quad \text{for } 1 \leq i < j \leq n.
\]
\end{definition}

For details, see \cite{kassel2008braid}.

In 1994, Long introduced a method for constructing braid group representations via an iterative process \cite{long1994constructing}. Let $F_n$ be a free group of rank $n$ and let  \( \text{Aut}(F_n) \) be the automorphism group of $F_n$.  Consider the Artin representation $\theta : B_n \longrightarrow \operatorname{Aut} (F_n)$ induced by the left action of the braid group $B_n$ on $\pi_1(D_n) = F_n$. 

\begin{definition}[Artin representation]
Let $x_{1},\ldots,x_{n}$ be the generators of $F_{n}$. The left action $\theta$ of $B_n$ on $F_n$ is defined as follows:

$$
\begin{array}{cccc}
\theta \colon& B_n & \longrightarrow & \operatorname{Aut}(F_n) \\
&\rotatebox{90}{$\in$} & & \rotatebox{90}{$\in$} \\
&\sigma_i& \longmapsto & \theta_{\sigma_i} 
\end{array}
$$

$$
\theta_{\sigma_i}(x_j) :=\left\{
\begin{array}{ll}
x_{i+1} & j=i \\
x_{i+1}^{-1}x_i x_{i+1} & j=i+1\\
x_j& j\neq i, i+1\\
\end{array}
\right.
.$$

\end{definition}

\begin{definition}[Semidirect product $F_n \rtimes_{\theta} B_n$]

The semidirect product of $B_n$ and $F_n$ with respect to $\theta$ is the group denoted by $F_n \rtimes_{\theta} B_n$, defined as follows:

\begin{description}
    \item[1. ] The underlying set of $F_n \rtimes_{\theta} B_n$ is the Cartesian product $ F_n \times B_n$.
    \item[2. ]The product is given by:
$$(h_1, g_1) \cdot (h_2, g_2) = (h_1 \cdot \theta_{g_1}(h_2), g_1 \cdot g_2),$$
where $h_1, h_2 \in F_n$, $g_1, g_2 \in B_n$, and $\theta_{g_1}(h_2)$ denotes the action of $g_1$ on $h_2$ via $\theta$.
\end{description}
\end{definition}

Hereinafter, we abbreviate the notation $\rtimes_{\theta}$ as simply $\rtimes$.

The Long-Moody construction is a method for constructing a representation of $B_n$ from any representation of $F_n \rtimes_\theta B_n$. It is a significant research subject from the perspectives of representation theory, the theory of linear differential equations in the complex domain, and knot theory.

 First, from the viewpoint of representation theory, the Long-Moody construction generalizes the Burau representation \cite{Burau1936} derived from the Alexander polynomial of knots. Subsequent research has revealed that this construction yields important representations of braid groups \cite{bigelow2008generalized}, notably unitary representations. Consequently, the Long-Moody construction serves as a unifying framework for classifying various representations of braid groups. An open problem of particular interest is whether every unitary representation of a braid group can be obtained via the construction \cite{birman2005braids}. Furthermore, Soulié \cite{soulie2019long, soulie2022generalized} has extended the construction by generalizing the braid group actions from Artin representations to Wada representations \cite{wada1992group, ITO20131754}. Aside from the Long-Moody construction, braid group representations can also be derived through generalizations of the Tong-Yang-Ma representations \cite{tong1996new}.

Second, we discuss the viewpoint of linear differential equations in complex domains. The action of elements of the fundamental group of the domain on the solution space of a differential equation is known as the monodromy representation. An important example of differential equations whose domain has a (pure) braid group as its fundamental group is given by the $n$-variable Knizhnik-Zamolodchikov (KZ-type) equations. A significant study by Drinfeld, Kanie, Kohno, and Tsuchiya establishes the connection between monodromy representations of KZ equations and representations of braid groups \cite{drinfeld1989quasi, Tsuchiya:1988fy, Kohno1987}. It is also known that the Lawrence-Krammer-Bigelow representations \cite{lawrence1990homological}, which are obtained by the construction, are related to the monodromy representations of KZ equations.

Third, we discuss the viewpoint of knot theory. Since any link can be expressed as the closure of a braid \cite{kassel2008braid}, the study of braid groups contributes substantially to knot invariants. In particular, the Burau representation recovers the Alexander polynomial, and it is known that twisted Alexander polynomials of knots can be obtained through the construction \cite{takano2024long}.

In our previous paper with Hiroe \cite{hiroe2023long}, we generalized the Long-Moody construction and obtained infinite sequences of braid group representations via the Katz-Long-Moody construction. The Katz-Long-Moody construction unifies the Long-Moody construction with the algorithm, based on twisted homology theory, for constructing local systems on $\mathbb{C}\setminus \lbrace n \text{ points}\rbrace$ introduced by Katz \cite{katz1996}. Katz's foundational theory of rigid local systems was extended by Dettweiler and Reiter \cite{DETTWEILER20071}, who developed a method for reconstructing Fuchsian linear differential equations with finitely many singularities. Haraoka further extended this and constructed a method for reconstructing $n$-variable KZ-type equations \cite{haraoka2012middle, haraoka2020multiplicative}. The KZ equation originally arose in conformal field theory as a differential equation for \( n \)-point correlation functions and has since emerged as a central object of study \cite{knizhnik1984}. Solutions to the KZ equation, including Selberg-type integrals, are closely related to various special functions such as the Appell-Lauricella hypergeometric series. KZ-type equations are a generalization of the KZ equation. 

In this paper, we establish a correspondence between the algebraic method of the Katz-Long-Moody construction and the analytic method of the middle convolution for KZ-type equations. In Section 2, we provide an explicit formulation of the Katz-Long-Moody construction in terms of matrix representations. In Section 3, we interpret Dettweiler-Reiter's method and Haraoka's method as approaches to constructing representations of the free group $F_n$ and the pure braid group $P_n$, respectively. In Section 4, we define a correspondence that connects these two methods and show that irreducibility is preserved under the construction. In Section 5, we show that the construction preserves unitarity of representations: if we take any unitary representation of $F_n \rtimes B_n$, then the Katz-Long-Moody construction yields another unitary representation. In this paper, we define unitarity as follows: 

\begin{definition}[Unitarity of representation \cite{long1994constructing}]

Let $G$ be a group and let $V$ be a finite-dimensional $\mathbb{C}$-vector space.

A linear representation $\rho\colon G \longrightarrow \mathrm{GL}(V)$ is said to be \emph{unitary relative to $H$} if there exists a non-degenerate Hermitian matrix $H$ such that $\rho(g)^{\dag}H\rho(g)=H$ holds for any $g \in G$, where $\dag$ denotes the conjugate transpose.  
\end{definition}

Unitary representations of braid groups play important roles in knot
invariants \cite{wenzl1993braids} and in topological quantum computation,
where braiding non-Abelian anyons induces unitary transformations on the
computational state space
\cite{nayak2008nonabelian, delaney2016local}.

In Section 5, we also propose an algorithm that determines the signature of the Hermitian matrix for all but finitely many parameters $\lambda$ with $|\lambda|=1$ and $\lambda\neq 1$.
 
\section{Algebraic construction of representations of the braid group}
\subsection{Settings for the algebraic construction}
Notably, in \cite{haraoka2020multiplicative}, Haraoka defined the paths of analytic continuation following the convention of Katz's theory. When considering the correspondence with the twisted Long-Moody construction, the geometric framework for defining the path plays a crucial role. Although the algebraic representation of the generators of the braid group remains the same, the definition of the half-twist, which serves as a generator of the mapping class group, admits two possible conventions.  
Specifically, it depends on whether $\sigma_i$ denotes a clockwise or counterclockwise rotation. Here, we adopt the convention that a counterclockwise rotation corresponds to the generator $\sigma_i$ of the braid group. Then, the generators $x_i$ of $F_n$ and the geometric realization of the Artin representation are defined in a manner consistent with this definition of the generators of $B_n$. The generator $x_i$ of the free group is represented as a path that loops clockwise around the point $p_i$.

\begin{itemize}
    \item generators of $B_n$: counterclockwise
    \item generators of $F_n$: clockwise
\end{itemize}

\begin{figure}[h]
    \centering
    \includegraphics[width=0.8\linewidth]{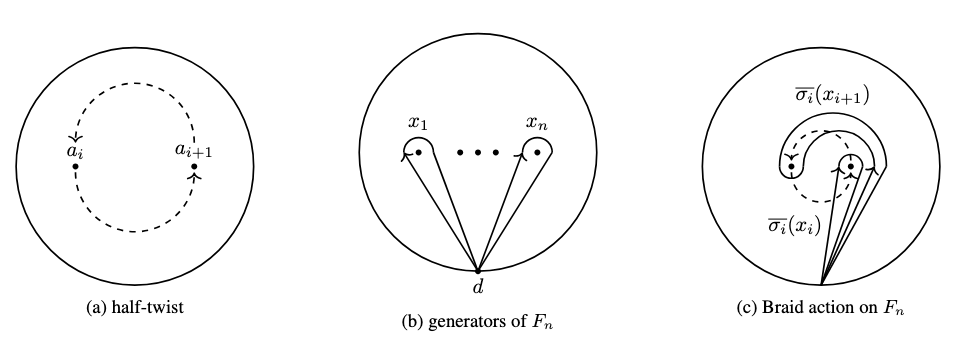}
    \caption{Geometric realization of Artin representation}
    \label{fig:geo-artin}
\end{figure}

The Artin representation corresponds to how the generators of the free group change when an element $\sigma_i$ of the braid group is given. When the action of the braid group is defined by a half-twist, the elements of the free group transform in accordance with the Artin representation. Hereafter, we denote the field by $\bm{k}$.

 \subsection{The Long-Moody construction}

\begin{definition}[Long-Moody construction]

Let $V$ be a finite-dimensional $\bm{k}$-vector space, and let $x_j, \sigma_i$ ($j=1, \dots, n$, $i=1, \dots, n-1$) be the generators of $F_n \rtimes B_n$. Given a group homomorphism

$$\rho \colon F_n \rtimes B_n \longrightarrow \mathrm{GL}(V),$$
we obtain the following group homomorphism 

$$\rho^{\mathrm{LM}} \colon B_n \longrightarrow \mathrm{GL}(V^{\oplus n})$$

where we denote
$$g_i :=\rho (x_i ),\ s_i :=\rho (\sigma_i),$$
$$\rho^{\mathrm{LM}}(\sigma_i):=s_i^{\oplus n}\cdot \left(\begin{array}{ccc}
I_{N(i-1)}&&\\
&R_i&\\
&&I_{N(n-i-1)}
\end{array}\right),$$
$$R_i := \left(\begin{array}{cc}
 0 & g_i \\
 I_N & I_N - g_{i+1}   \\
\end{array}\right).$$

\end{definition}

Bigelow extended this construction to subgroups of $B_n$ and obtained the following result.

\begin{theorem}[Long-Moody construction\cite{bigelow2008generalized}]
Let $V$ be a finite-dimensional $\bm{k}$-vector space, and let $B$ be any subgroup of $B_n$. Applying the LM construction to a group homomorphism

$$\rho \colon F_n \rtimes B \longrightarrow \mathrm{GL}(V),$$ we obtain a homomorphism of $B$,

$$\rho^{\mathrm{LM}}\ |_B \colon B \longrightarrow \mathrm{GL} (V^{\oplus n}).$$ This homomorphism is called the LM construction for the subgroup $B$. \end{theorem} 

For the proof of the theorem, see \cite{bigelow2008generalized}.

\subsection{The Katz-Long-Moody construction}

Various representations can be obtained by the LM construction; however, the structure of $F_n$ is lost in the process. To address this, we obtain new representations while preserving the information of $F_n$ by using the convolution approach of Dettweiler and Reiter \cite{DETTWEILER20071}. In our previous paper \cite{hiroe2023long}, we generalized the Long-Moody construction in this way and named it the twisted Long-Moody construction.

\begin{definition}[Dettweiler-Reiter's convolution \cite{DETTWEILER2000761}]

Let $V$ be an $N$-dimensional vector space over $\bm{k}$.
Let \(\lambda \in \bm{k}^{\times}\), and let \(x_i\) be the generators of \(F_n\). For any $\rho\colon F_n \longrightarrow \mathrm{GL}(V)$, we write $g_i := \rho(x_i)$ and abbreviate $I_N$ as $1$.

$$\rho^{DR}_{\lambda}\colon F_n \longrightarrow \mathrm{GL}(V^{\oplus n})$$

\begin{align*}
\rho^{DR}_{\lambda}(x_i)
:=\left(\begin{array}{ccccccc}
1&&&&&&\\
&\ddots&&&&&\\
&&1&&&&\\
\lambda(g_1-1)&\cdots&\lambda(g_{i-1}-1)&\lambda g_i&g_{i+1}-1&\cdots&g_n-1\\
&&&&1&&\\
&&&&&\ddots&\\
&&&&&&1\\
\end{array}\right).
\end{align*}

Hereinafter, we abbreviate the identity matrix $I_k$ of size $k$ simply as $1$ when it is clear from the context.

We define the following notation. Let $E_{ij}$ be the $n\times n$ block matrix, with blocks of size $N\times N$, whose $(i,j)$-th block is $I_N$ and whose other blocks are $O_N$. Let $A\circ E_{ij}$ be the block matrix whose $(i,j)$-th block is $A$ and whose other blocks are $O_N$. By this notation, $\rho^{DR}_{\lambda}(x_i)
$ is written as $I_{Nn} + \Sigma_{j=1}^{n} \lambda^{\star ij}(g_j-\lambda_{\star ij})\circ E_{ij}$. Here we denote 
$$
\begin{array}{cl}
\lambda^{\star ij}&= \left\{
\begin{array}{ll}
\lambda & (i \geq j)\\
1 & (i < j)
\end{array}
\right.
\end{array} \text{, and}
\begin{array}{cl}
\lambda_{\star ij}&= \left\{
\begin{array}{ll}
\lambda^{-1} & (i = j)\\
1 & (i \neq j)
\end{array}
\right.
\end{array}.
$$

\end{definition}

Using Dettweiler-Reiter's convolution, the twisted Long-Moody construction is defined as follows.

\begin{definition}[Twisted Long-Moody construction]\normalfont
\label{tLM}

Let $V$  be an $N$-dimensional vector space over the field $\bm{k}$.

Let the generators of $F_n$ and $B_n$ be $x_j, \sigma_i, j=1,\cdots, n, i=1, \cdots, n-1$.
In addition, let $B \subseteq B_n$ and $\lambda \in \bm{k}^{\times}$. For a group homomorphism $$\rho\colon F_n \rtimes B \longrightarrow \mathrm{GL}(V),$$ we construct a representation
$$\rho^{LM}_{\lambda}\colon F_n \rtimes B \longrightarrow \mathrm{GL}(V^{\oplus n})$$

on the generators as follows:

\begin{align*}
\rho^{LM}_{\lambda}(x_i):=& \rho^{DR}_{\lambda}(x_i)  \\
\rho^{LM}_{\lambda}(\sigma_i):=& \rho^{LM}(\sigma_i) 
\end{align*}

\end{definition}

\begin{proposition}\label{prop:tLM-welldef}
The assignment in Definition~\ref{tLM} is well defined; that is, it extends to a group homomorphism $\rho^{LM}_{\lambda}\colon F_n \rtimes B \longrightarrow \mathrm{GL}(V^{\oplus n})$.
\end{proposition}

\begin{proof}
This has already been shown in our previous paper \cite{hiroe2023long}; however, we give an algebraic proof here. For the generators $x_j, \sigma_i$ ($j=1,\cdots, n$, $i=1, \cdots, n-1$) of $F_n$ and $B_n$, we set $g_i:= \rho(x_i)$ and $s_i:= \rho(\sigma_i)$. It suffices to verify the Artin relations:

$$
\rho^{LM}_{\lambda}(\sigma_i x_j) =\left\{
\begin{array}{ll}
\rho^{LM}_{\lambda}(x_{i+1}\sigma_i) & j=i \\
\rho^{LM}_{\lambda}(x_{i+1}^{-1}x_i x_{i+1}\sigma_i) & j=i+1\\
\rho^{LM}_{\lambda}(x_j\sigma_i)& j\neq i, i+1.\\
\end{array}
\right.
$$

We introduce the following notation:
$\widetilde{G_i} = \rho^{LM}_{\lambda}(x_{i}) - I_{Nn}$ and $\widetilde{S_i} = (s_i^{-1})^{\oplus n}\rho^{LM}_{\lambda}(\sigma_i) - I_{Nn}$. That is,
\[
\widetilde{G_i}
= \sum_{k=1}^{n} \lambda^{\star ik}(g_k-\lambda_{\star ik}I)\circ E_{ik}.
\]
Here $\theta_{\sigma_i}$ denotes the action of $\sigma_i$ on $F_n$, given by
$\theta_{\sigma_i}(x_i)=x_{i+1}$, $\theta_{\sigma_i}(x_{i+1})=x_{i+1}^{-1}x_i x_{i+1}$,
and $\theta_{\sigma_i}(x_k)=x_k$ for $k\neq i,i+1$. Since $\rho$ is a homomorphism on
$F_n\rtimes B_n$, we have $\rho(\theta_{\sigma_i}^{-1}(x_k)) = s_i^{-1}g_k\, s_i$ for the corresponding blocks.

We denote $(\widetilde{G_j})_{\Theta_i} := \bigl(s_i^{\oplus n}\bigr)^{-1}\widetilde{G_j}\, s_i^{\oplus n}$, so that $\widetilde{G_j}\cdot s_i^{\oplus n} = s_i^{\oplus n}\cdot (\widetilde{G_j})_{\Theta_i}$. Componentwise, this conjugation replaces each $g_k$ by $\rho(\theta_{\sigma_i}^{-1}(x_k))$; for instance,
$$
(\widetilde{G_{i+1}})_{\Theta_i}
= \sum_{k=1}^{n} \lambda^{\star\, i+1,k}\bigl(\rho(\theta_{\sigma_i}^{-1}(x_k))-\lambda_{\star\, i+1,k}\bigr)\circ E_{i+1,k}.
$$

For $j=i$,
$$
    \begin{array}{cl}
    \rho^{LM}_{\lambda}(\sigma_i x_i) 
    =&s_i^{\oplus n}\cdot (I_{Nn} + \widetilde{S_i})(I_{Nn} + \widetilde{G_i})
        =s_i^{\oplus n}\cdot(I_{Nn} + \widetilde{S_i} + \widetilde{G_i} + \widetilde{S_i}\widetilde{G_i})\\
    \rho^{LM}_{\lambda}(x_{i+1}\sigma_i ) 
    =& (I_{Nn}+ \widetilde{G_{i+1}})s_i^{\oplus n}(I_{Nn}+\widetilde{S_i})=s_i^{\oplus n}(I_{Nn}+ (\widetilde{G_{i+1}})_{\Theta_i})(I_{Nn}+\widetilde{S_i}).
\end{array}
$$
Then, since $\rho^{LM}_{\lambda}(\sigma_i x_i) - \rho^{LM}_{\lambda}(x_{i+1}\sigma_i ) = \widetilde{G_i} - (\widetilde{G_{i+1}})_{\Theta_i} + \widetilde{S_i}\widetilde{G_i} - (\widetilde{G_{i+1}})_{\Theta_i}\widetilde{S_i} =O$, $\rho^{LM}_{\lambda}(\sigma_i x_i) = \rho^{LM}_{\lambda}(x_{i+1}\sigma_i ) $ holds.\\

For $j=i+1$, 
\begin{equation}
\begin{array}{ll}
     &\rho^{LM}_{\lambda}(x_{i+1}\sigma_i x_{i+1})-\rho^{LM}_{\lambda}(x_{i}x_{i+1}\sigma_i )
     \\=&(I_{Nn} + \widetilde{G_{i+1}})s_i^{\oplus n}(I_{Nn}+\widetilde{S_i})(I_{Nn}+ \widetilde{G_{i+1}}) - (I_{Nn} + \widetilde{G_{i}})(I_{Nn}+ \widetilde{G_{i+1}})s_i^{\oplus n}(I_{Nn}+\widetilde{S_i})\\
     =&(I_{Nn}+\widetilde{G_{i+1}})s_i^{\oplus n}(\widetilde{G_{i+1}}-(\widetilde{G_{i+1}})_{\Theta_i}+\widetilde{S_i}\widetilde{G_{i+1}}-(\widetilde{G_{i+1}})_{\Theta_i}\widetilde{S_i})\\
     =&O.
\end{array}
\end{equation}

For $j\neq i, i+1$, since $\theta_{\sigma_i}(x_j)=x_j$ we have $(\widetilde{G_j})_{\Theta_i}=\widetilde{G_j}$, and therefore
$
\begin{array}{ll}
&\rho^{LM}_{\lambda}(\sigma_i x_j) -\rho^{LM}_{\lambda}(x_j\sigma_i ) \\
=& s_i^{\oplus n}(I_{Nn} + \widetilde{S_i})(I_{Nn} + \widetilde{G_j}) -(I_{Nn} + \widetilde{G_j}) s_i^{\oplus n}(I_{Nn} + \widetilde{S_i})\\
=& s_i^{\oplus n}(\widetilde{G_j} - (\widetilde{G_j})_{\Theta_i} + \widetilde{S_i}\widetilde{G_j} - (\widetilde{G_j})_{\Theta_i}\widetilde{S_i})\\
=&s_i^{\oplus n}(\widetilde{G_j} - \widetilde{G_j} + \widetilde{S_i}\widetilde{G_j} - \widetilde{G_j}\widetilde{S_i})\\
=&O.
\end{array}$

\end{proof}

The Katz-Long-Moody construction is given by identifying a $\rho^{LM}_{\lambda}$-invariant subspace $V_{\text{inv}}$ of $V^{\oplus n}$ and considering the induced action on the corresponding quotient space $V^{\oplus n}\slash V_{\text{inv}}$. Here we define the subspaces $K$ and $L$ of $V^{\oplus n}$ as follows, where $G_j := \rho^{LM}_{\lambda}(x_j)$.

$$K := \left\lbrace 
\begin{pmatrix}
 w_1 \\ \vdots \\ w_n 
\end{pmatrix}; w_j \in \mathrm{Ker} (g_j -1) (1\leq j \leq n)
\right\rbrace,
$$

$$\begin{displaystyle}
L := \bigcap_{j=1}^{n} \mathrm{Ker} (G_j -I_{Nn}).
\end{displaystyle}
$$

\begin{lemma}
    $K + L $ is $\rho^{LM}_{\lambda}$-invariant.
\end{lemma}

This lemma has already been proven in our previous research \cite{hiroe2023long}; here we give a proof in terms of concrete matrix representations.

\begin{proof}   
We show the invariance in several steps.

For $K$, it suffices to show that $\rho^{LM}_{\lambda}(x_j)K \subset K$ and $\rho^{LM}_{\lambda}(\sigma_i)K \subset K.$ 

Take any element in $K$, $\begin{pmatrix}
 w_1 \\ \vdots \\ w_n 
\end{pmatrix}; w_j \in \mathrm{Ker}(g_{j} -1)\ (1\leq j \leq n).
$

$$\rho^{LM}_{\lambda}(\sigma_i)\cdot
\begin{pmatrix}
 w_1 \\ \vdots \\ w_n 
\end{pmatrix}
=
\begin{pmatrix}
s_i w_1 \\
\vdots\\
s_i g_{i} w_{i+1}\\
s_i w_i + s_i (1-g_{i+1}) w_{i+1}\\
\vdots\\
s_i w_n 
\end{pmatrix}
=
\begin{pmatrix}
s_i w_1 \\
\vdots\\
s_i g_{i} w_{i+1}\\
s_i w_{i}\\
\vdots\\
s_i w_n 
\end{pmatrix}\\
=
\begin{pmatrix}
s_i w_1 \\
\vdots\\
g_{i+1} s_i w_{i+1}\\
s_i w_{i}\\
\vdots\\
s_i w_n 
\end{pmatrix}
$$

For $k \neq i, i+1$, $(s_i w_k) = s_i g_k w_k = g_k (s_i w_k)$. So $s_i w_k \in \operatorname{Ker}(g_k-1).$

For $k=i$, $g_i (s_i g_{i} w_{i+1})  = s_i g_i g_{i+1} w_{i+1} = s_i g_i w_{i+1}$. So, $s_i g_{i} w_{i+1} \in \operatorname{Ker}(g_i -1)$.

For $k=i+1$, $g_{i+1} s_i w_i = s_i g_i w_i = s_i w_i$. So $(g_{i+1}-1)s_i w_i=O$, thus $s_i w_i \in \operatorname{Ker}(g_{i+1}-1)$.

$$
\begin{array}{cl}
&\rho^{LM}_{\lambda}(x_j)\cdot
\begin{pmatrix}
 w_1 \\ \vdots \\ w_n 
\end{pmatrix}
=\begin{pmatrix}
w_1 \\
\vdots\\
w_{j}^{'}\\
\vdots\\
w_n 
\end{pmatrix}
=
\begin{pmatrix}
w_1 \\
\vdots\\
\lambda g_j w_j\\
\vdots\\
w_n 
\end{pmatrix}
=
\begin{pmatrix}
w_1 \\
\vdots\\
\lambda w_j\\
\vdots\\
w_n 
\end{pmatrix}
\end{array}.
$$
Here we denote $\lambda(g_1 -1)w_1 + \cdots + \lambda(g_{j-1} -1)w_{j-1} +\lambda g_{j}w_{j} +(g_{j+1}-1)w_{j+1} +\cdots + (g_{n}-1)w_{n}$ as $w_j^{'}$.

So, 
$$
\rho^{LM}_{\lambda}(x_j)\cdot
\begin{pmatrix}
 w_1 \\ \vdots \\ w_n 
\end{pmatrix}
\in K.
$$

For $L$, observe that
$L = \bigcap_{j=1}^{n} \mathrm{Ker}\bigl(\rho^{LM}_{\lambda}(x_j) - I_{Nn}\bigr)$
is exactly the subspace of $F_n$-fixed vectors of $V^{\oplus n}$. Since $F_n$
is a normal subgroup of $F_n \rtimes B_n$, for any
$\beta \in F_n \rtimes B_n$ and any $x \in F_n$ we have
$\beta^{-1} x \beta \in F_n$. Hence, for $\bm{w} \in L$,
$$
\rho^{LM}_{\lambda}(x)\, \rho^{LM}_{\lambda}(\beta)\, \bm{w}
= \rho^{LM}_{\lambda}(\beta)\, \rho^{LM}_{\lambda}(\beta^{-1} x \beta)\, \bm{w}
= \rho^{LM}_{\lambda}(\beta)\, \bm{w},
$$
so that $\rho^{LM}_{\lambda}(\beta)\, \bm{w} \in L$. Therefore $L$ is invariant
under the whole of $\rho^{LM}_{\lambda}$; in particular
$\rho^{LM}_{\lambda}(x_j) L \subseteq L$ and
$\rho^{LM}_{\lambda}(\sigma_i) L \subseteq L$.

\end{proof} 
\begin{definition}[the Katz-Long-Moody construction \cite{hiroe2023long}]\normalfont
The action of $\rho^{LM}_{\lambda}$ descends to the quotient space $V^{\oplus n}/(K + L)$. We define the Katz-Long-Moody construction $\rho^{KLM}_{\lambda}$ as the induced action of $\rho^{LM}_{\lambda}$ on $V^{\oplus n}/(K + L)$.

\end{definition}

\section{Analytic construction of monodromy representations of KZ-type equations}

\subsection{KZ type equation}
To establish a correspondence between the algebraic construction and the analytical construction, it is necessary to understand the differences in the settings related to their respective generators and relations. In each construction method, the various settings and their geometric realizations are described first, followed by a discussion on the construction of representations of the braid group. 

The analytical method discussed here, known as the multiplicative middle convolution \cite{haraoka2012middle, haraoka2020multiplicative}, defines $B_n$ and its action on $F_n$ as follows. For the generators of $B_n$, the half-twist is defined counterclockwise, as in the algebraic construction. The generators of $F_n$ correspond to the paths of analytic continuation, but unlike the algebraic construction, the paths are defined counterclockwise. Then, each path $\alpha_i$ changes as follows in accordance with the action of the braid group.

The Artin representation corresponds to how the generators of the free group change when an element $\sigma_i$ of the braid group is given. When the action of the braid group is defined by a half-twist, the elements of the free group transform in accordance with the Artin representation.

The generators of the pure braid group are defined as follows:

$$\widetilde{\sigma_{ij}} := (\widetilde{\sigma_i}\cdots\widetilde{\sigma_{j-2}})\widetilde{\sigma_{j-1}}^{2}(\widetilde{\sigma_i}\cdots\widetilde{\sigma_{j-2}})^{-1}$$

 In this study, we focus on the pure braid group $P_n$, which is a subgroup of the braid group, in order to establish a relationship with complex analysis.

\begin{definition}[Pure braid group $P_n$]

Let $S_n$ be the symmetric group on $n$ letters.

$$P_n := \operatorname{Ker}(\Pi\colon B_n \ni \sigma_i \mapsto (i, i+1)\in S_n  )$$

\end{definition}

The generators of $P_n$ are given by $\sigma_{ij} := \sigma_i \cdots \sigma_{j-2} (\sigma_{j-1})^2 (\sigma_i \cdots \sigma_{j-2})^{-1}$.

The following standard decomposition of the pure braid group is well known
(see, for example, \cite{FadellNeuwirth1962,Birman1974}): letting $F_n$ denote
the free group of rank $n$,
\[
P_{n+1} \cong F_n \rtimes P_n.
\]

Let the generators of $P_{n+1}$ be $\sigma_{ij}$, $0\leq i < j \leq n$, and let the generators of $F_n \rtimes P_n$ be $x_j, j = 1,\dots, n$ and $\sigma_{ij}, 1\leq i < j \leq n$.

Then, 
$$
\begin{array}{cccc}
f\colon&P_{n+1} & {\longrightarrow} & F_n \rtimes P_n \\
&\rotatebox{90}{$\in$} & & \rotatebox{90}{$\in$} \\
&\sigma_{ij} & \longmapsto & 
\left\{
\begin{array}{ll}
x_j & i=0 \\
\sigma_{ij} & i > 0
\end{array}
\right.
\end{array}
$$
is an example of the group isomorphism above. 

 \subsection{Settings for the analytic construction}

The monodromy representation is an (anti-)representation of the fundamental group of the domain of a differential equation. A linear transformation of the fundamental solution matrix is determined by analytic continuation along the paths corresponding to the generators of the fundamental group of the domain. Depending on whether the action of the fundamental group on the solution space is taken as a left action or a right action, it becomes either an anti-representation or a representation.

For $i =0, 1, \dots, n$, let $a_i$ be a point in $\mathbb{R}$, such that $  a_0<\dots < a_n.$\ 
$Q_{n+1}:=\lbrace a_0, \dots, a_n \rbrace,\ 
{Q_{n+1}}^{i}:=Q_{n+1}\backslash\{a_i\}$, and
\begin{align*}
Q^{n+1}&:=\left\{ (z_0, \dots, z_n) \in \mathbb{C}^{n+1}\ \middle| \ \prod_{i<j}(z_j - z_i)=0\right\}
.\end{align*}

\begin{definition}[KZ type equation]
Let $n$ and $N$ be positive integers and assume that $ z=(z_0, \cdots, z_n)\in \mathbb{C}^{n+1}$. A KZ-type equation is a system of linear partial differential equations

\begin{equation}\label{eq:KZ}
    \frac{\partial u}{\partial z_i} = \sum_{j\neq i} \frac{A_{ij}}{z_i-z_j}u,\quad A_{ij} = A_{ji},\quad i=0, \dots,n
\end{equation}

where $A_{i,j}$ are $N\times N$ constant matrices. Note that $\mathbb{C}^{n+1}  \backslash Q^{n+1}$ is the domain of the KZ-type equation.

\end{definition}

In addition, we assume the following integrability conditions:

\begin{align*}
\lbrack A_{i,j},\ A_{k,l}\rbrack=&O\ \ \{i, j\}\cap \{k, l\} = \emptyset\\
\lbrack A_{i,j},\ A_{i,k}+A_{j,k}\rbrack=&O\ \ i\neq j, i\neq k, j\neq k
\end{align*}

Here, following previous studies \cite{haraoka2020multiplicative}, we assume that the monodromy representation is an anti-representation. Let $Q$ be a tuple $(a_0, \ldots, a_n)$. 

\begin{definition}[Generators of $\pi_1(\mathbb{C}^{n+1}\backslash Q^{n+1}, Q)$, $\lbrack\alpha_{ij}\rbrack$]
A path on $\mathbb{C}^{n+1}$, $[\alpha_{ij}]$ is defined below.

$$
\begin{array}{cccc}
\lbrack \alpha_{ij}\rbrack \colon  &[0, 1]  & \stackrel{}{\longrightarrow} & \mathbb{C}^{n+1}\backslash Q^{n+1}\\
&\rotatebox{90}{$\in$} & & \rotatebox{90}{$\in$} \\
&t & \longmapsto & (a_0, \dots, a_{i-1}, {\gamma_i}^j(t), a_{i+1}, \dots, a_n)
\end{array}
$$

Here, ${\gamma_i}^j$ is a simple closed curve with base point $a_i$ that encircles only $a_j$ among the points of $Q_{n+1}$. Suppose that
$$
\begin{array}{cccc}
{\gamma_i}^j \colon &[0, 1]& \longrightarrow &\mathbb{C}\backslash{Q_{n+1}}^i\\
&\rotatebox{90}{$\in$} & & \rotatebox{90}{$\in$} \\
&t & \longmapsto & {\gamma_i}^j(t)\\
\end{array}$$

$${\gamma_i}^j(0) ={\gamma_i}^j(1)=a_i.$$

\end{definition}

The homotopy classes of the paths $\lbrack\alpha_{ij}\rbrack$ are known to generate $\pi_1(\mathbb{C}^{n+1}\backslash Q^{n+1}, Q)$. The map sending $\lbrack\alpha_{ij}\rbrack$ to $\sigma_{ij}$ is a group homomorphism from $\pi_1(\mathbb{C}^{n+1}\backslash Q^{n+1}, Q)$ to $P_{n+1}$.

Let $\mathcal{U}$ be a fundamental matrix solution in the neighbourhood of $Q$, and let $\alpha_{ij*}\mathcal{U}$ denote analytic continuation of $\mathcal{U}$ along the path $\lbrack\alpha_{ij}\rbrack$. Then, there exist matrices $M_{ij} \in \mathrm{GL}(N, \mathbb{C})$ such that $\alpha_{ij*}\mathcal{U} = \mathcal{U} M_{ij}$. The matrices $M_{ij}$ are called the monodromy matrices for the loops $\alpha_{ij}$. Haraoka's convolution of KZ-type equations is a method for constructing a KZ-type equation whose coefficient matrices $B$ are constant complex matrices of size $Nn\times Nn$ from a KZ-type equation whose coefficient matrices $A$ are constant complex matrices of size $N\times N$. Haraoka proposed a method to construct a new tuple of monodromy matrices $(N_{ij})_{0\leq i < j \leq n}$ from a KZ-type equation whose monodromy matrices are $(M_{ij})_{0\leq i < j \leq n}$ through the convolution of the KZ-type equation \cite{haraoka2020multiplicative}. $N_{ij}$ is an $n\times n$ block matrix whose blocks are polynomials in the $M_{kl}$. Since the domain of the KZ-type equation obtained by the convolution coincides with that of the original equation, we can also consider analytic continuations along the same paths $\alpha_{ij}$. Denoting the analytic continuation by $\alpha_{ij*}$, the monodromy representation corresponding to the convolution of the KZ-type equation can be written as $\alpha_{ij*}\mathcal{V} = \mathcal{V} N_{ij}$, where $\mathcal{V}$ is a fundamental matrix solution. The fundamental group of the domain of the KZ-type equation is the fundamental group of the configuration space of $n+1$ ordered points, which is isomorphic to the pure braid group $P_{n+1}$. Henceforth, we can identify $\lbrack\alpha_{ij}\rbrack$ with $\sigma_{ij}$.

\begin{figure}[h]
    \centering
    \includegraphics[width=1\linewidth]{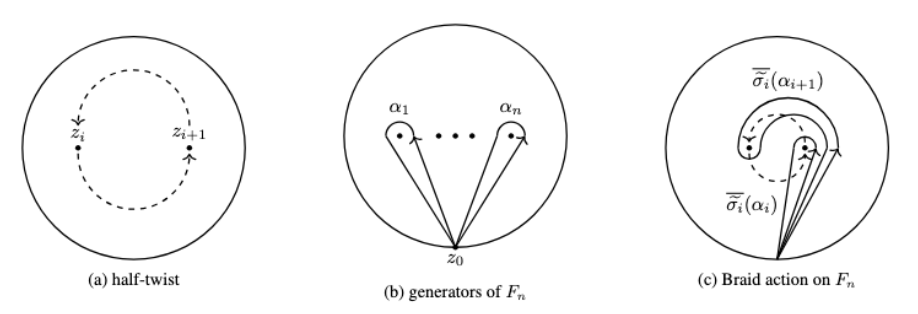}
    \caption{Settings for analytic construction}
    \label{fig:setting-analytic}
\end{figure}

\subsection{Multiplicative middle convolution}

In order to formulate the multiplicative middle convolution for KZ-type equations, we first recall the definition of the additive middle convolution for KZ-type equations~\cite{haraoka2020multiplicative}. We consider the additive middle convolution of a KZ-type equation in the $z_0$-direction. Take a fundamental matrix solution $\mathcal{U}(z_0, z_1, \dots, z_n)$. As a function of the single variable $z_0$, $\mathcal{U}$ satisfies the ordinary differential equation
\begin{equation}
    \frac{\partial u}{\partial z_0} = \left(\sum_{j=1}^n \frac{A_{0j}}{z_0 - z_j}\right) u,
    \label{eq:3.1}
\end{equation}
in $z_0$. This is the restriction of \eqref{eq:KZ} in the $z_0$-direction. The fundamental matrix solution of this equation is given as follows.

Let $l \in \mathbb{C}$ be a parameter, and set $\lambda := e^{2\pi i l}$;
thus $\lambda$ is the multiplicative counterpart of the additive parameter $l$,
namely the factor by which the branch of $(t - z_0)^{l}$ is multiplied along a
loop encircling $t = z_0$ once positively. Define the matrix function
\begin{equation}
    \mathcal{V}(z_0, z_1, \dots, z_n) = \left( \int_{\Delta_k} \frac{\mathcal{U}(t, z_1, \dots, z_n)}{t - z_j}(t - z_0)^{l} dt \right)_{1 \leq j,\  k \leq n},
    \label{eq:3.2}
\end{equation}
where $\Delta_k$ ($1 \leq k \leq n$) are defined as follows.

Take an auxiliary point $b\in\mathbb{R}$ such that
$z_0<b<z_1$.
Let $\delta_0$ be a path from $b$ to $z_0$, and for
$1\le k\le n$, let $\delta_k$ be a path from $z_{k-1}$
to $z_k$.
Let $\alpha_0$ be a loop based at $b$ encircling $z_0$
once positively, and for $1\le k\le n$, let $\alpha_k$
be a loop based at $b$ encircling $z_k$ once positively
via the lower half-plane.
Define the Pochhammer cycle by
\[
\Delta_k=[\alpha_k,\alpha_0]
\qquad (1\le k\le n).
\]
In the module of twisted chains associated with the local system determined by
$(t-z_0)^{l}\,\mathcal{U}$, the Pochhammer cycle admits the regularized
expression
\[
\Delta_k=(\delta_1+\cdots+\delta_k)(I_N-M_{0k})(1-\lambda)
\qquad (1\le k\le n),
\]
where the composite path $\delta_1+\cdots+\delta_k$ from $b$ to $z_k$ is loaded
with the indicated matrix coefficient (see \cite{haraoka2020multiplicative}).
In particular, the two displays describe the same twisted cycle $\Delta_k$.

By the linearity of the integral, this solution space can be regarded as a vector space with the integration paths $\Delta_1, \dots, \Delta_n$ serving as its basis. We will specify them in the next section. It is shown that $\mathcal{V}$ satisfies the ordinary differential equation
\begin{equation}
    \frac{\partial v}{\partial z_0} = \left(\sum_{j=1}^n \frac{B_{0j}}{z_0 - z_j}\right) v,
    \label{eq:3.3}
\end{equation}
in $z_0$, where $B_{0j}$ ($1 \leq j \leq n$) are constant matrices of size $nN$ given by
\begin{equation}
    B_{0j} = \begin{pmatrix}
        0 & \cdots & 0 & \cdots & 0 \\
        \vdots & \ddots & \vdots & \ddots & \vdots \\
        A_{01} & \cdots & A_{0j} + lI_N & \cdots & A_{0n} \\
        \vdots & \ddots & \vdots & \ddots & \vdots \\
        0 & \cdots & 0 & \cdots & 0
    \end{pmatrix}.
    \label{eq:3.4}
\end{equation}

Equation \eqref{eq:3.3} is called the convolution equation of \eqref{eq:KZ} with parameter $\lambda$. Haraoka showed in \cite{haraoka2012middle, haraoka2020multiplicative} that the ordinary differential equation can be prolonged to a Pfaffian system
\begin{equation}
    d\nu = \left( \sum_{0 \leq i < j \leq n} B_{ij} d\log(z_i - z_j) \right) \nu,
    \label{eq:3.5}
\end{equation}
in $(z_0, z_1, \dots, z_n)$ with constant matrices $B_{ij}$ which are uniquely determined. We call the system \eqref{eq:3.5} the convolution system of \eqref{eq:3.1} in the $z_0$-direction with parameter $\lambda$.

The basis of the solution space can be chosen as follows. 
$$(\Delta_1 \dots \Delta_n) :=(\delta_1 \dots\delta_n)P$$

Here, $P$ is defined as follows to represent the twisted cycles.
\begin{equation}
    P = \begin{pmatrix}
        (1-\lambda)(1-M_{01}) &(1-\lambda)(1-M_{02}) &\dots& (1-\lambda)(1-M_{0n})\\
                              &(1-\lambda)(1-M_{02}) &  &\vdots\\
                              &                      & \ddots &\vdots\\
                              &                      &  & (1-\lambda)(1-M_{0n})
    \end{pmatrix}
\end{equation}

Then, the monodromy matrices $N_{ij}$ with respect to this basis take the following form:

\begin{center}
$
\begin{array}{cc}
(\sigma_{ij*}\Delta_1 \dots \sigma_{ij*}\Delta_n) &=(\sigma_{ij*}\delta_1 \dots\sigma_{ij*}\delta_n) P\\
&= (\delta_1 \dots\delta_n) X(\sigma_{ij})P\\
&= (\delta_1 \dots\delta_n) P N_{ij}\\
&= (\Delta_1 \dots\Delta_n) N_{ij}     
\end{array}
$
\end{center}

Here, $X(\sigma_{ij})$ satisfies $(\sigma_{ij*}\delta_1 \dots\sigma_{ij*}\delta_n) 
= (\delta_1 \dots\delta_n) X(\sigma_{ij}).$

\begin{theorem}[Haraoka \cite{haraoka2020multiplicative} Theorem 5.2]

Let $n$ and $N$ be positive integers, and let $V$ be an $N$-dimensional vector space over $\mathbb{C}$. We assume that $\lambda \in \mathbb{C}^{\times} $.

For the following anti-homomorphism, with generators $\widetilde{\sigma_{ij}}$ $(0\leq i < j \leq n)$ of $P_{n+1}$,

$$
\begin{array}{cccc}
\widetilde{\rho}\colon &P_{n+1} & \stackrel{}{\longrightarrow} & \text{GL}(V) \\
&\rotatebox{90}{$\in$} & & \rotatebox{90}{$\in$} \\
&\widetilde{\sigma_{ij}} & \longmapsto & M_{ij}
\end{array}
,$$
we obtain the following new anti-homomorphism $\widetilde{\rho}^{H}_{\lambda}$:
$$
\begin{array}{cccc}
\widetilde{\rho}^{H}_{\lambda} \colon &P_{n+1} & \stackrel{}{\longrightarrow} & \text{GL}(V^{\oplus n}) \\
&\rotatebox{90}{$\in$} & & \rotatebox{90}{$\in$} \\
&\widetilde{\sigma_{ij}}& \longmapsto & N_{ij}
\end{array}
$$
where $N_{ij}$ is defined as follows.

\begin{description}
\item[For $i=0$]

$$
    N_{0j}:= I_{Nn} + \Sigma^{n}_{k=1}
    \lambda^{\star jk}(M_{0k}-\lambda_{\star jk}) \circ E_{jk}
$$

\item[For $i>0$]

$$
    N_{ij}:=\left(
    \begin{array}{ccc}
    M_{ij}^{\oplus(i-1)}&&\\
    &N_{ij}^{'}&\\
    &&M_{ij}^{\oplus(n-j)}
    \end{array}
    \right).
$$

Here, for $k=i+1, \dots, j-1$, we denote $N_{ij}^{'}$ as \\
$
    \left(\begin{array}{ccccc}
    M_{0j}M_{ij}& X_{i+1}&\cdots&X_{j-1}&M_{0j}M_{ij}(1-M_{0j})\\
    &M_{0, i}^{-1}M_{ij}M_{0j}&&&\\
    &&\ddots&&\\
    &&&M_{0, i}^{-1}M_{ij}M_{0j}&\\
    M_{ij}(1-M_{0, i})& Y_{i+1}&\cdots&Y_{j-1}&M[i, j, 0]\\
    \end{array}\right),$\\

and $
\begin{array}{cc}
    X_k \ =\  &(M_{0j}M_{ij}-M_{0, i}^{-1}M_{ij}M_{0j})(1-M_{0k}),\\
    Y_k \ =\  &M_{ij}(1-M_{0, i})(1-M_{0k}),\\
    M[i, j, 0]\ =\ &M_{ij}-M_{ij}M_{0j}+M_{0j}M_{ij}M_{0, i}.    
\end{array}
$  
\end{description}
\end{theorem}

Haraoka's convolution is defined analytically, but it can also be defined over any field $\bm{k}$ as a method of constructing a new anti-representation $(N_{ij})_{0\leq i < j \leq n}$ of $P_{n+1}$ from the anti-representation $(M_{ij})_{0\leq i < j \leq n}$ of $P_{n+1}$. Therefore, we define a new method of constructing anti-representations of $P_{n+1}$ over any field $\bm{k}$ and call it Haraoka's convolution.

\begin{definition}[Haraoka's convolution]

Let $n$ and $N$ be positive integers, and let $V$ be an $N$-dimensional vector space over $\bm{k}$. In addition, we assume that $\lambda \in \bm{k}^{\times}$.
Then, for the following anti-homomorphism with the generators 
$(\widetilde{\sigma_{ij}})_{0\leq i < j \leq n}$ of $P_{n+1}$,

$$
\begin{array}{cccc}
\rho\colon &P_{n+1} & \stackrel{}{\longrightarrow} & \text{GL}(V) \\
&\rotatebox{90}{$\in$} & & \rotatebox{90}{$\in$} \\
&\widetilde{\sigma_{ij}} & \longmapsto & M_{ij}
\end{array}
,$$
we define the anti-homomorphism $\widetilde{\rho}^{\mathrm{H}}_{\lambda}$ of $P_{n+1}$ as follows and call it Haraoka's convolution.
$$
\begin{array}{cccc}
\widetilde{\rho}^{\mathrm{H}}_{\lambda} \colon &P_{n+1} & \stackrel{}{\longrightarrow} & \text{GL}(V^{\oplus n}) \\
&\rotatebox{90}{$\in$} & & \rotatebox{90}{$\in$} \\
&\widetilde{\sigma_{ij}}& \longmapsto & N_{ij}
\end{array}
$$
\end{definition}
See \cite{haraoka2020multiplicative}.

The action of $\widetilde{\rho^{H}_{\lambda}}$ descends to the quotient space $\mathbb{C}^{Nn}/(K + L)$. We call the induced action on $\mathbb{C}^{Nn}/(K + L)$ the multiplicative middle convolution.

It is shown in \cite{DETTWEILER2000761} that, under some generic conditions, the resulting representation is irreducible if the original representation is irreducible.

\section{The Katz--Long--Moody construction and the multiplicative middle convolution}
\subsection{The Haraoka-Long correspondence}
The first main result is an anti-isomorphism between the restriction of the twisted LM construction to $P_{n+1}$ and Haraoka's convolution. Henceforth, we identify \(P_{n+1}\) with \(F_n \rtimes P_n\) as follows.

$$
\begin{array}{cccc}
 &P_{n+1} & \stackrel{}{\longrightarrow} & F_{n}\rtimes P_{n}\\
&\rotatebox{90}{$\in$} & & \rotatebox{90}{$\in$} \\
&\sigma_{ij} & \longmapsto &\left\{
\begin{array}{ll}
x_{j}& i=0 \\
\sigma_{i j} & i>0
\end{array}
\right.
\end{array}
$$

With this identification, the twisted LM construction for $P_{n+1}$ can be rewritten as follows.

$$
\begin{array}{ccccll}
\rho \colon  &F_n \rtimes P_{n}  & \stackrel{}{\longrightarrow} & \mathrm{GL}(V)&\\
&\rotatebox{90}{$\in$} & & \rotatebox{90}{$\in$} &\\
&x_{j} & \longmapsto & \rho(\sigma_{0j}) & \\
&\sigma_{ij} & \longmapsto & \rho(\sigma_{ij}) &i > 0
\end{array}
$$ we rewrite this as 
$$
\rho^{\mathrm{LM}}_{\lambda}(\sigma_{ij}):=\left\{
\begin{array}{ll}
\rho^{\mathrm{DR}}_{\lambda}(x_j)& i=0 \\
\rho^{\mathrm{LM}}_{\lambda}(\sigma_{ij}) & i>0
\end{array}
\right.
$$

Note that the twisted LM construction is a way of constructing representations of $P_{n+1}$, while Haraoka's convolution is a way of constructing anti-representations of $P_{n+1}$. To construct the correspondence between them, we define anti-isomorphisms of groups as follows.

\begin{definition}[group anti-isomorphism $\psi^{op}$]
For a group $G$, we define anti-isomorphism $\mathrm{inv}$ as $\mathrm{inv} \colon G \ni g \mapsto g^{-1} \in G$.

For a group isomorphism $\psi \colon G \longrightarrow H$,
we define anti-isomorphism $\psi^{op} \colon G \longrightarrow H$ as 
$$\psi^{op} := \psi \circ \mathrm{inv}.$$ In the same manner,\\
for anti-isomorphism $\eta^{\mathrm{anti}} \colon G \longrightarrow H$, we define isomorphism $(\eta^{\mathrm{anti}})^{op}\colon G \longrightarrow H$ as 
$$(\eta^{\mathrm{anti}})^{op} := \eta^{\mathrm{anti}} \circ \mathrm{inv}.$$
\end{definition}

\begin{lemma}\label{generator}
For the generators $(\sigma_{i})$ of $B_n$, we define $\widetilde{\sigma_i}:= \sigma_i^{-1}$.

Here, we assume that
\begin{align*}
    \sigma_{ij}&:= \sigma_{i}\cdots \sigma_{j-2}(\sigma_{j-1})^2(\sigma_{j-2}^{-1}\cdots \sigma_{i}^{-1})\\
\widetilde{\sigma_{ij}}&:= \widetilde{\sigma_{i}}\cdots \widetilde{\sigma_{j-2}}(\widetilde{\sigma_{j-1}})^{2}(\widetilde{\sigma_{j-2}}^{-1}\cdots \widetilde{\sigma_{i}}^{-1})
,\end{align*}
then \((\sigma_{ij})\) and \((\tilde\sigma_{ij})\) both generate \(P_{n+1}\).

Moreover, the following relation holds:

$$\rho^{op}(\widetilde{\sigma_{ij}})= \rho(\sigma_{ij}),$$
$$
\begin{array}{cccccc}
\rho^{op} \colon &P_{n+1} & \stackrel{\mathrm{inv}}{\longrightarrow} & P_{n+1} &\stackrel{\rho}{\longrightarrow} & \mathrm{GL}(V) \\
&\rotatebox{90}{$\in$} & & \rotatebox{90}{$\in$}& &\rotatebox{90}{$\in$}\\
&{\widetilde{\sigma_{ij}}} & \longmapsto & \sigma_{i j} &\longmapsto&\rho(\sigma_{ij})
\end{array}
$$.
\end{lemma}

\begin{theorem}[the Haraoka-Long correspondence]\label{haraoka-long}
For a group homomorphism $\rho\colon P_{n+1} \longrightarrow \mathrm{GL}(V)$, let $\rho_\lambda^{\mathrm{LM}}$ be a generalized LM of $\rho$ for the generator  $(\sigma_{ij})$,
let $(\rho^{op})_\lambda^{\mathrm{H}}$ be Haraoka's convolution of $\rho^{op}$ for the generator $(\widetilde{\sigma_{ij}})$.
Then we have 
$$(\rho_\lambda^{\mathrm{LM}})^{op} = (\rho^{op})_\lambda^{\mathrm{H}}.$$

$$
\begin{array}{ccc}
\rho &\longrightarrow & \rho^{LM}_{\lambda} \\
\rotatebox{90}{$\longleftarrow$} &\circlearrowright & \rotatebox{90}{$\longleftarrow$}\\
\rho^{\operatorname{op}}&\longrightarrow& (\rho^{LM}_{\lambda})^{\operatorname{op}} 
\end{array}
$$

\end{theorem}

\begin{proof}

When we have $\rho(\sigma_{ij})=\rho^{op}(\widetilde{\sigma_{ij}})$, we will prove that $(\rho_\lambda^{\mathrm{LM}})^{op}(\widetilde{\sigma_{ij}}) = (\rho^{op})_\lambda^{\mathrm{H}}(\widetilde{\sigma_{ij}})$.
Here we divide the argument into the following two cases:
(1)\ $i=0$, and (2)\ $i>0$. We define the symbols as follows:
$\rho^{op}(\widetilde{\sigma_{ij}})=M_{ij}$, and $ (\rho^{op})_\lambda^{\mathrm{H}}(\widetilde{\sigma_{ij}})=N_{ij}$.

\begin{description}
    \item[(1)\ For $i=0$]
\end{description}    
It follows immediately from the definitions of $\rho^{LM}_\lambda$, $(\rho^{op})^{\mathrm{H}}_\lambda$, together with Lemma \ref{generator} that 
\begin{align*}
    &(\rho^{op})_\lambda^{\mathrm{H}}
(\widetilde{\sigma_{0j}})\\
    =&
    \left(\begin{array}{ccccccc}
    1&&&&&&\\
    &\ddots&&&&&\\
    &&1&&&&\\
    \lambda(M_{01}-1)&\cdots&\lambda(M_{0 j-1}-1)&\lambda M_{0j}&(M_{0, j+1}-1)&\cdots&(M_{0n}-1)\\
    &&&&1&&\\
    &&&&&\ddots&\\
    &&&&&&1
    \end{array}\right)\\
    =&(\rho_\lambda^{\mathrm{LM}})(\sigma_{0j})=(\rho_\lambda^{\mathrm{LM}})^{op}(\widetilde{\sigma_{0j}}).
\end{align*}

The last equality follows from Lemma~\ref{generator}
and the identification $\sigma_{0j}\leftrightarrow x_j$.

\begin{description}
    \item[(2)\ For $i>0$]
\end{description}    

    Fix $n>2$. 
    By computing the matrix entries and using mathematical induction on $i$, we show that 
    
    $$(\rho_\lambda^{\mathrm{LM}})^{op}(\widetilde{\sigma_{ij}}) = (\rho^{op})_\lambda^{\mathrm{H}}
(\widetilde{\sigma_{ij}}).$$
    
    That is, we show that 
    \begin{enumerate}
        \item For each $i$, we have $(\rho_\lambda^{\mathrm{LM}})^{op}(\widetilde{\sigma_{i, i+1}}) = (\rho^{op})_{\lambda}^{\mathrm{H}}(\widetilde{\sigma_{i, i+1}})$.
        \item For any $i$ and $j$, assume that $(\rho_\lambda^{\mathrm{LM}})^{op}(\widetilde{\sigma_{i+1, j}}) = (\rho^{op})_{\lambda}^{\mathrm{H}}(\widetilde{\sigma_{i+1, j}})$. Then it follows that $(\rho_\lambda^{\mathrm{LM}})^{op}(\widetilde{\sigma_{i, j}}) = (\rho^{op})_{\lambda}^{\mathrm{H}}(\widetilde{\sigma_{i, j}})$.
    \end{enumerate}

1. The proof is based on computing the entries of the matrices.

$$
\begin{array}{cl}
&(\rho^{\mathrm{LM}}_{\lambda})^{op}(\widetilde{\sigma_{i, i+1}})=\rho^{\mathrm{LM}}_{\lambda}(\sigma_i^2) 
\\
&=s_i^{\oplus n}\cdot \left(\begin{array}{ccc}
I_{N(i-1)}&&\\
&R_i&\\
&&I_{N(n-i-1)}
\end{array}\right)\cdot s_i^{\oplus n}\cdot \left(\begin{array}{ccc}
I_{N(i-1)}&&\\
&R_i&\\
&&I_{N(n-i-1)}
\end{array}\right)\\
&=(s_i^2)^{\oplus n}\cdot\left(\begin{array}{ccc}
I_{N(i-1)}&&\\
&\Theta_i(R_i)&\\
&&I_{N(n-i-1)}
\end{array}\right)\cdot \left(\begin{array}{ccc}
I_{N(i-1)}&&\\
&R_i&\\
&&I_{N(n-i-1)}
\end{array}\right)\\
&=(s_i^2)^{\oplus n}\cdot\left(\begin{array}{ccc}
I_{N(i-1)}&&\\
&\Theta_i(R_i)\cdot R_i&\\
&&I_{N(n-i-1)}
\end{array}\right)\\
&=
\left(\begin{array}{ccc}
\rho(\sigma_i^2)^{\oplus (i-1)}&&\\
&\rho(\sigma_i^2)^{\oplus 2}\cdot\Theta_i(R_i)\cdot R_i&\\
&&\rho(\sigma_i^2)^{\oplus (n-i-1)}
\end{array}\right)\\
&=
\left(\begin{array}{ccc}
\rho^{op}(\widetilde{\sigma_i}^2)^{\oplus (i-1)}&&\\
&\rho(\sigma_i^2)^{\oplus 2}\cdot\Theta_i(R_i)\cdot R_i&\\
&&\rho^{op}(\widetilde{\sigma_i}^2)^{\oplus (n-i-1)}
\end{array}\right)\\
&=
\left(\begin{array}{ccc}
{M_{i, i+1}}^{\oplus (i-1)}&&\\
&\rho(\sigma_i^2)^{\oplus 2}\cdot\Theta_i(R_i)\cdot R_i&\\
&&{M_{i, i+1}}^{\oplus (n-i-1)}
\end{array}\right).
\end{array}
$$

For brevity, write
\[
\theta_i(g_k):=\rho\bigl(\theta_{\sigma_i}^{-1}(x_k)\bigr)=s_i^{-1}g_k s_i.
\]

$$
\begin{array}{cl}
&\rho(\sigma_i^2)^{\oplus 2}\cdot\Theta_i(R_i)\cdot R_i \\
=& 
\rho(\sigma_i^2)^{\oplus 2}\cdot
\begin{pmatrix}
0 & \theta_i(g_i) \\
1 & 1-\theta_i(g_{i+1})\\
\end{pmatrix}\cdot
\begin{pmatrix}
0 & g_i \\
1 & 1-g_{i+1}\\
\end{pmatrix}
 \\
    =& \rho(\sigma_i^2)^{\oplus 2}\cdot\begin{pmatrix}
\theta_i(g_i) & \theta_i(g_i)(1-g_{i+1}) \\
1-\theta_i(g_{i+1}) & g_i + (1-\theta_i(g_{i+1}))(1-g_{i+1})\\
\end{pmatrix}\\
    =&\begin{pmatrix}
 \rho(x_{i+1})\cdot\rho(\sigma_i^2)&  \rho(x_{i+1})\rho(\sigma_i^2)(1-\rho(x_{i+1})) \\
 \rho(\sigma_i^2)\cdot(1-\rho(x_{i})) &  \rho(\sigma_i^2)\cdot(1- \rho(x_{i+1})+\rho(x_i)\rho(x_{i+1}) )\\
\end{pmatrix}\\
    =&\begin{pmatrix}
 M_{0,i+1}M_{i, i+1} &  M_{0, i+1}M_{i, i+1}(1-M_{0,i+1}) \\
 M_{i, i+1}(1-M_{0, i}) &  M[i, i+1, 0]\\
\end{pmatrix}
\end{array}.
$$

Then, $(\rho^{\mathrm{LM}}_{\lambda})^{op}(\widetilde{\sigma_{i, i+1}}) = N_{i, i+1}=(\rho^{op})_\lambda^{\mathrm{H}}(\widetilde{\sigma_{i, i+1}})$ holds.
\\
2. For $1\leq i< i+1<j\leq n$, we assume that $(\rho^{op})_\lambda^{\mathrm{H}}(\widetilde{\sigma_{i+1, j}})=(\rho^{\mathrm{LM}}_{\lambda})^{op}(\widetilde{\sigma_{i+1, j}}) $. Under this assumption, it suffices to show that $(\rho^{op})_\lambda^{\mathrm{H}}(\widetilde{\sigma_{i, j}})=(\rho^{\mathrm{LM}}_{\lambda})^{op}(\widetilde{\sigma_{i, j}}) $.
$$
\begin{array}{cl}
    RHS=(\rho^{\mathrm{LM}}_{\lambda})^{op}(\widetilde{\sigma_{i, j}}) 
&=(\rho^{\mathrm{LM}}_{\lambda})^{op}(\widetilde{\sigma_{i}^{-1}})
    \cdot(\rho^{\mathrm{LM}}_{\lambda})^{op}(\widetilde{\sigma_{i+1, j}})
\cdot(\rho^{\mathrm{LM}}_{\lambda})^{op}(\widetilde{\sigma_{i}})\\
&=(\rho^{\mathrm{LM}}_{\lambda})(\sigma_{i}^{-1})
\cdot(\rho^{op})_\lambda^{\mathrm{H}}{(\widetilde{\sigma_{i+1, j}})}
\cdot(\rho^{\mathrm{LM}}_{\lambda})(\sigma_{i})
\end{array}
$$
Therefore, it remains to show that 
\begin{align*}
\rho^{\mathrm{LM}}_{\lambda}(\sigma_i^{-1})
N_{i+1,j}
\rho^{\mathrm{LM}}_{\lambda}(\sigma_i)
=
N_{i,j}.\end{align*}

\[
\begin{aligned}{ll}
&\rho^{\mathrm{LM}}_{\lambda}(\sigma_{i}^{-1})
\cdot N_{i+1, j}
\cdot\rho^{\mathrm{LM}}_{\lambda}(\sigma_{i})  \\
     =&
     \begin{pmatrix}
         (s_i^{-1})^{\oplus (i-1)}&&\\
    &R_i^{-1}    (s_i^{-1})^{\oplus 2}&\\
    &&    (s_i^{-1})^{\oplus (n-i-1)}
     \end{pmatrix}
     \cdot
     N_{i+1, j}\cdot
     \begin{pmatrix}
s_i^{\oplus (i-1)}&&\\
    &s_i^{\oplus 2}R_i&\\
    && s_i^{\oplus (n-i-1)}
     \end{pmatrix}\\
=&
     \begin{pmatrix}
         (s_i^{-1})^{\oplus (i-1)}&&\\
    &R_i^{-1}    (s_i^{-1})^{\oplus 2}&\\
    &&    (s_i^{-1})^{\oplus (n-i-1)}
     \end{pmatrix}
     \cdot
     \left(
    \begin{array}{ccc}
    (M_{i+1, j})^{\oplus i}&&\\
    &N_{i+1, j}^{'}&\\
    &&M_{i+1, j}^{\oplus(n-j)}
    \end{array}
    \right)\\
    &     \cdot
     \begin{pmatrix}
s_i^{\oplus (i-1)}&&\\
    &s_i^{\oplus 2}R_i&\\
    && s_i^{\oplus (n-i-1)}
     \end{pmatrix}\\
=&
\begin{pmatrix}
         (s_i^{-1}M_{i+1, j}s_i)^{\oplus (i-1)}&&\\
    &X&\\
    &&    (s_i^{-1}M_{i+1, j}s_i)^{\oplus (n-j)}
     \end{pmatrix}\\
=&
\begin{pmatrix}
         {M_{ij}}^{\oplus (i-1)}&&\\
    &X&\\
    &&   {M_{ij}}^{\oplus (n-j)}
     \end{pmatrix}=N_{i,j}.     
\end{aligned}
\]

Here we compute $X$ as follows.

$
\begin{array}{rl}
&X=
\begin{pmatrix}
    R_i^{-1}\cdot(s_i^{-1})^{\oplus 2}&\\
    &(s_i^{-1})^{\oplus (j-i-2)}
\end{pmatrix}
\begin{pmatrix}
    M_{i+1, j}&\\
    &N_{i+1, j}^{'}
\end{pmatrix}
\begin{pmatrix}
        s_i^{\oplus 2}\cdot R_i&\\
    &s_i^{\oplus (j-i-2)}
\end{pmatrix}\\
     =&\begin{pmatrix}
    R_i^{-1}\cdot(s_i^{-1})^{\oplus 2}&\\
    &(s_i^{-1})^{\oplus (j-i-2)}
\end{pmatrix}
\begin{pmatrix}
    M_{i+1, j}&\\
    &N_{i+1, j}^{'}
\end{pmatrix}
\begin{pmatrix}
        s_i^{\oplus 2}\cdot R_i&\\
    &s_i^{\oplus (j-i-2)}
\end{pmatrix}\\
=&\begin{pmatrix}
    (g_{i+1}-1)g_i^{-1}s_i^{-1}&s_i^{-1}&\\
    g_i^{-1}&0&\\
    &&(s_i^{-1})^{\oplus (j-i-2)}
\end{pmatrix}
\cdot
\begin{pmatrix}
    M_{i+1, j}&\\
    &N_{i+1, j}^{'}
\end{pmatrix}
\\
\cdot&
\begin{pmatrix}
    O&s_i g_i&\\
    s_i&s_i(1-g_{i+1})&O\\
    &&s_i^{\oplus (j-i-2)}\\
\end{pmatrix}\\
&=N_{ij}^{'}
\end{array}
$

By the preceding argument,

    \begin{description}
        \item[1.\ ] for any $i$, $(\rho_\lambda^{\mathrm{LM}})^{op}(\widetilde{\sigma_{i, i+1}}) = (\rho^{op})_{\lambda}^{\mathrm{H}}(\widetilde{\sigma_{i, i+1}})$,
        \item[2.\ ] for any $i+1, j$, if $(\rho_\lambda^{\mathrm{LM}})^{op}(\widetilde{\sigma_{i+1, j}}) = (\rho^{op})_{\lambda}^{\mathrm{H}}(\widetilde{\sigma_{i+1, j}})$ is true, then $(\rho_\lambda^{\mathrm{LM}})^{op}(\widetilde{\sigma_{i, j}}) = (\rho^{op})_{\lambda}^{\mathrm{H}}(\widetilde{\sigma_{i, j}})$.
\end{description}
This proves case (2) by induction on \(i\).

Finally, by (1), (2), it follows that 
$$(\rho_\lambda^{\mathrm{LM}})^{op} = (\rho^{op})_\lambda^{\mathrm{H}}.$$

\end{proof}

Since $K+L$ is invariant under $\rho^{\mathrm{LM}}_\lambda$,
Theorem~\ref{haraoka-long} induces the following correspondence
on the quotient. Hence, we obtain the following corollary.

\begin{corollary}

The Haraoka-Long correspondence induces a correspondence between the Katz-Long-Moody construction (restricted to $P_{n+1}$) and the multiplicative middle convolution.

\end{corollary}
 \subsection{Correspondence with the two settings}

Here we discuss the differences in the settings between the algebraic construction method (the KLM construction) and the analytic construction method (the multiplicative middle convolution), which are essential for establishing the correspondence between them.

The KLM construction yields representations, i.e., group homomorphisms, whereas the multiplicative middle convolution produces anti-representations, i.e., group anti-homomorphisms. Moreover, we note that the orientations of paths defining the generators of the fundamental groups of the domains differ between the two methods. This difference in path orientation causes a change in the generators of the pure braid group.

 \subsection{Irreducibility}

In \cite{haraoka2020multiplicative}, Theorem 5.7, it is shown that the multiplicative middle convolution produces an irreducible representation under the assumption that the tuple $(M_{0j})_{1\leq j\leq n}$ is irreducible, that is, the matrices $M_{01}, \dots, M_{0n}$ admit no common nontrivial invariant subspace. By combining this theorem with the Haraoka-Long correspondence, we obtain an irreducible representation via the Katz-Long-Moody construction, provided that the restriction $\rho|_{F_n}$ is irreducible, that is, the tuple $(g_1, \dots, g_n)$ admits no common nontrivial invariant subspace. Note that this hypothesis is in general stronger than the irreducibility of $\rho$ as a representation of $F_n \rtimes B_n$.

\section{Unitarity}
\subsection{Monodromy invariant Hermitian form}

In this section, we discuss the unitarity of the braid group representations,
which is our second result. Monodromy-invariant Hermitian forms have long played an important role
in the study of Fuchsian differential equations. In particular, Haraoka
explicitly constructed such a form for the Pochhammer equation and used
its definiteness to study finite monodromy \cite{haraoka1994}.
The behavior of invariant Hermitian forms under middle convolution has
also been studied from Hodge-theoretic and low-rank viewpoints
\cite{dettweiler2013hodge, adachi2024unitary}.
Our result extends these considerations to the multiplicative middle
convolution for KZ-type equations of general rank and provides a recursive
algorithm for computing the signature of the resulting Hermitian form.

Throughout this section, we fix $\bm{k} = \mathbb{C}$, assume
$\lambda \in \mathbb{C}$ with $|\lambda| = 1$ and $\lambda \neq 1$
(for $\lambda = 1$ the multiplicative middle convolution reduces, under
suitable non-degeneracy conditions, to the identity operation, cf.\
\cite{haraoka2020multiplicative}),
fix a square root $\lambda^{\frac{1}{2}}$ of $\lambda$, and denote by
$X^{\dag}$ the adjoint matrix of $X$.
 
Recall that a linear representation
$\rho\colon G \longrightarrow \mathrm{GL}(V)$ is unitary relative to a
non-degenerate Hermitian matrix $H$ if
$\rho(g)^{\dag} H \rho(g) = H$ holds for any $g \in G$
(see the definition in the introduction).

The prototype of such a result on the braid group side is Squier's theorem
\cite{squier1984burau}: the reduced Burau representation is unitary relative
to a Hermitian form defined over $\mathbb{Z}[t, t^{-1}]$ with the involution
$t \mapsto t^{-1}$; specializing $t$ to the unit circle yields a complex
Hermitian form, whose signature depends on the location of $t$.
 
In \cite{long1994constructing}, Long proved that if $\rho$ is unitary, then so
is $\rho^{LM}_{s}$ for some generic value $s$, following the method of
Deligne--Mostow \cite{delignemonodromy}. Here,
$\rho^{LM}_{s}(\sigma_i) := s \cdot \rho^{LM}(\sigma_i)$.
 
We extend this result and show that unitarity is preserved by the
Katz-Long-Moody construction under some conditions. It also follows that the
multiplicative middle convolution of KZ-type equations preserves unitarity.
First, as mentioned in the introduction, we construct a Hermitian matrix, not
necessarily non-degenerate, that satisfies the unitarity condition, and show
that it induces a non-degenerate Hermitian form on the quotient defining
$\rho^{KLM}_{\lambda}$. Finally, we give an algorithm computing the signature
of this Hermitian matrix.
 
Define
$$(\widetilde{H})_{jk} := h_{jk}
= \lambda^{jk}\, H\, (g_j^{-1} - \lambda_{jk} I)(g_k - 1),
\qquad 1 \leq j, k \leq n,$$
$$
\lambda^{jk}=\left\{
\begin{array}{ll}
\lambda^{-\frac{1}{2}} & j\leq k\\
\lambda^{\frac{1}{2}} & j> k
\end{array}
\right.,\qquad
\lambda_{jk}=\left\{
\begin{array}{ll}
\lambda & j=k\\
1 & \text{otherwise.}
\end{array}
\right.
$$
Note that replacing $\lambda^{\frac{1}{2}}$ by $-\lambda^{\frac{1}{2}}$
replaces $\widetilde{H}$ by $-\widetilde{H}$.
 
\begin{theorem}\label{thm:uni}
Assume that there exists a non-degenerate Hermitian matrix $H$ satisfying
$\rho(\beta)^{\dag} H \rho(\beta) = H$ for any $\beta \in F_n \rtimes B_n$.
Then $\widetilde{H}$ is Hermitian and satisfies the unitarity condition
$$
\rho^{LM}_{\lambda}(\beta)^{\dag}\, \widetilde{H}\, \rho^{LM}_{\lambda}(\beta)
= \widetilde{H}
\qquad \text{for any } \beta \in F_n \rtimes B_n.
$$
\end{theorem}
 
\begin{theorem}[Unitarity of $\rho^{KLM}_{\lambda}$]\label{thm:non-degenerate}
Under the assumptions of Theorem~\ref{thm:uni},
$$
\mathrm{Ker}\, \widetilde{H} = K + L.
$$
In particular, $\widetilde{H}$ induces a non-degenerate Hermitian form
$\widetilde{H}^{KLM}$ on $V^{\oplus n} / (K + L)$, and $\rho^{KLM}_{\lambda}$
is unitary relative to $\widetilde{H}^{KLM}$. Thus, if $\rho$ is unitary, then
so is $\rho^{KLM}_{\lambda}$.
\end{theorem}
 
\begin{proof}[Proof of Theorem~\ref{thm:uni}]
 
Write $\rho^{LM}_{\lambda}(x_i) = I + \widetilde{G_i}$, where
$\widetilde{G_i}$ has non-zero blocks only in the $i$-th block row, and denote
by $g_{ik}$ its $(i, k)$-th block:
$$g_{ik} := \lambda^{k} (g_k - \lambda_{k} I),
\qquad
\lambda^{k}=\left\{
\begin{array}{ll}
\lambda & k \leq i\\
1 & k > i
\end{array}
\right.,\qquad
\lambda_{k}=\left\{
\begin{array}{ll}
\lambda^{-1} & k=i\\
1 & \text{otherwise.}
\end{array}
\right.
$$
 
For $i = 1, \cdots, n$, we will show that
$$
\begin{array}{cc}
    &(I + \widetilde{G_i})^{\dag}\, \widetilde{H}\, (I + \widetilde{G_i})
    = \widetilde{H}\\
\iff& \widetilde{G_i}^{\dag} \widetilde{H}
    + \widetilde{H} \widetilde{G_i}
    + \widetilde{G_i}^{\dag} \widetilde{H} \widetilde{G_i} = O.\\
\end{array}
$$
 
Since $\widetilde{G_i}$ has non-zero blocks only in the $i$-th block row,
 
$$
\begin{array}{cl}
    (k, l)\text{-th block of (LHS)}
    &={\sum_{s=1}^n} g_{sk}^{\dag} h_{s, l}
    +\sum_{t=1}^n h_{k, t} g_{t, l}
    +\sum_{s=1}^{n} g_{sk}^{\dag} \left\lbrace\sum_{t=1}^n h_{s, t} g_{t, l}\right\rbrace\\
    &= g_{ik}^{\dag} h_{i, l}
    + h_{k, i} g_{i, l}
    + g_{ik}^{\dag} h_{i, i} g_{i, l}\\
    &= (h_{l, i}\, g_{ik})^{\dag}
    + h_{k, i} g_{i, l}
    + g_{ik}^{\dag} h_{i, i} g_{i, l},
\end{array}
$$
where in the last line we used $h_{il} = h_{li}^{\dag}$.
 
For each term, the following equality holds, where we repeatedly use
$g^{\dag} H = H g^{-1}$ and $g^{-\dag} H = H g$, which follow from
$g^{\dag} H g = H$.
$$
\begin{array}{rl}
    h_{k, i}\, g_{i, l}
    =&\lambda^{ki}\lambda^l H (g_k^{-1}-\lambda_{ki})(g_i-I)(g_l -\lambda_{l}),
\end{array}
$$
 
$$
\begin{array}{rl}
    (h_{l, i}\, g_{i, k})^{\dag}
    =&\left(\lambda^{li}\lambda^{k} H (g_l^{-1}-\lambda_{li})(g_i-I)(g_k-\lambda_k)\right)^{\dag}\\
    =&\overline{\lambda^{li}\lambda^{k}}\,
    (g_k^{\dag}-\overline{\lambda_k})
    (g_i^{\dag}-I)
    (g_l^{-\dag}-\overline{\lambda_{li}})\, H\\
    =&\overline{\lambda^{li}\lambda^{k}}\, H
    (g_k^{-1} -\overline{\lambda_k})
    (g_i^{-1}-I)
    (g_l-\overline{\lambda_{li}}),
\end{array}
$$
 
$$
\begin{array}{rl}
    g_{ik}^{\dag}\, h_{i, i}\, g_{i, l}
    =&\left(\lambda^k(g_k -\lambda_k I)\right)^{\dag}
    \lambda^{ii} H (g_i^{-1}-\lambda)(g_i-I)\,
    \lambda^{l}(g_l -\lambda_{l} I)\\
    =&\overline{\lambda^{k}}\lambda^{-\frac{1}{2}}\lambda^{l}
    (g_k^{\dag}-\overline{\lambda_k}) H
    (g_i^{-1}-\lambda)(g_i-I)
    (g_l -\lambda_{l} I)\\
    =&\overline{\lambda^{k}}\lambda^{-\frac{1}{2}}\lambda^{l}
    H
    (g_k^{-1}-\overline{\lambda_k})
    (g_i^{-1}-\lambda)(g_i-I)
    (g_l -\lambda_{l} I).
\end{array}
$$
 
Consequently,
 
$$
\begin{array}{cl}
    &(k, l)\text{-th block of (LHS)}=\\
    &
    \lambda^{l}             a_1 g_k^{-1}g_ig_l 
    + \overline{\lambda^k}   a_2g_k^{-1}g_i^{-1}g_l
    - \lambda^l\lambda_l     a_1 g_k^{-1}g_i 
    -\overline{\lambda^{k}}   a_3 g_k^{-1}g_i^{-1}\\
    &-\lbrace \lambda^l a_1 + \overline{\lambda^k}a_2 \rbrace g_k^{-1} g_l
    +\lbrace \overline{\lambda^k} a_3 + \lambda^l \lambda_l a_1 \rbrace g_k^{-1}
    - \lambda^l a_4 g_ig_l
    - \overline{\lambda_k }\overline{\lambda^k} a_2 g_i^{-1}g_l\\
    &+ \lambda^l\lambda_l a_4 g_i
    + \overline{\lambda^k}\overline{\lambda_k} a_3 g_i^{-1}
    - \lbrace \overline{\lambda^k}\overline{\lambda_k} a_3 + \lambda^l\lambda_l a_4 \rbrace,
\end{array}
$$
 
where $a_1 = \lambda^{ki}- \lambda^{\frac{1}{2}}\overline{\lambda^k}, a_2 = \overline{\lambda^{li}}-\lambda^{-\frac{1}{2}}\lambda^l, a_3 = \overline{\lambda^{li}} \overline{\lambda_{li}}-\lambda^{ii}\lambda^l \lambda_l, a_4=\lambda^{ki}\lambda_{ki}-\lambda^{\frac{1}{2}}\overline{\lambda^k}\overline{\lambda_k}.$ 
 
Note that $a_1$ and $a_4$ depend only on $k$ (through $\lambda^{ki}$, $\lambda^k$, $\lambda_{ki}$, $\lambda_k$), while $a_2$ and $a_3$ depend only on $l$ (through $\lambda^{li}$, $\lambda^l$, $\lambda_{li}$, $\lambda_l$; note also $\lambda^{ii}$, which depends only on $i$). Hence it suffices to verify $a_1=a_4=0$ separately for each of $k<i$, $k=i$, $k>i$, and $a_2=a_3=0$ separately for each of $l<i$, $l=i$, $l>i$; together these combinations exhaust all cases for the pair $(k,l)$.
 
For $k < i$ and $k = i$,
$a_1 = a_4 = \lambda^{-\frac{1}{2}} - \lambda^{\frac{1}{2}}\lambda^{-1} = 0$
and
$a_1 = a_4 = \lambda^{-\frac{1}{2}}\lambda - \lambda^{\frac{1}{2}}\lambda^{-1}\lambda = 0$,
respectively (for $a_4$ at $k = i$, note $\lambda_{ki} = \lambda$ and
$\overline{\lambda_k} = \lambda$).
For $k > i$,
$a_1 = a_4 = \lambda^{\frac{1}{2}} - \lambda^{\frac{1}{2}} = 0$.
 
For $l < i$ and $l = i$,
$a_2 = a_3 = \lambda^{\frac{1}{2}} - \lambda^{-\frac{1}{2}}\lambda = 0$
(for $a_3$ at $l = i$, note
$\overline{\lambda_{li}} = \lambda^{-1}$ and $\lambda_l = \lambda^{-1}$, so
$a_3 = \lambda^{\frac{1}{2}}\lambda^{-1} - \lambda^{-\frac{1}{2}}\lambda\lambda^{-1} = 0$).
For $l > i$,
$a_2 = a_3 = \lambda^{-\frac{1}{2}} - \lambda^{-\frac{1}{2}} = 0$.
 
Thus, the $(k,l)$-th block is identically zero.
 
For $S_i := \rho^{LM}_{\lambda}(\sigma_i)$, $i=1,\cdots, n-1$, writing
$s_i := \rho(\sigma_i)$, we will show that
$$
    S_i^{\dag}\, \widetilde{H}\, S_i = \widetilde{H}.
$$

As in the case of $\rho^{LM}_{\lambda}(x_i)$, both sides are Hermitian, so it
suffices to compute the $(k,l)$-blocks listed below; the remaining blocks are
obtained by taking adjoints. Note that the computations for
$k \neq i, i+1$ are valid for every such $k$, regardless of the relative order
of $k$ and $l$.

If $k, l \neq i, i+1$,
 
$$
\begin{array}{cl}
(\operatorname{LHS})_{kl}&= s_i^{\dag}
(\lambda^{kl}H (g_k^{-1}-1)(g_l-1))
s_i\\
&=\lambda^{kl}H s_i^{-1}s_i (g_k^{-1}-1)(g_l-1)\\
&=(\widetilde{H})_{kl}.
\end{array}
$$
 
If $k \neq i, i+1$ and $l=i$, 
 
$$
\begin{array}{cl}
&(\operatorname{LHS})_{ki}\\
=& s_i^{\dag}
(\lambda^{ki}H (g_k^{-1}-1)(g_i-1))
s_i
-\lambda^{k, i+1}H s_i^{-1}(g_k^{-1}-1)(g_i-1)s_i\\
&+\lambda^{k, i+1}H s_i^{-1}(g_k^{-1}-1)(g_{i+1}-1)s_i\\
=&\lambda^{k, i}H(g_k^{-1}-1)(g_i-1)\\
=&(\widetilde{H})_{ki}.
\end{array}
$$
 
If $k \neq i, i+1$ and $l=i+1$, 
 
$$
\begin{array}{cl}
&(\operatorname{LHS})_{k, i+1}\\=& s_i^{\dag}
(\lambda^{ki}H (g_k^{-1}-1)(g_{i+1}-1))
s_i
+\lambda^{k, i}H s_i^{-1}(g_k^{-1}-1)(g_i-1)s_i g_i\\
&-\lambda^{k, i+1}H s_i^{-1}(g_k^{-1}-1)(g_{i+1}-1)s_i g_{i+1}\\
=&\lambda^{k, i}H(g_k^{-1}-1)(g_{i+1}-1)\\
=&(\widetilde{H})_{k, i+1}.
\end{array}
$$
 
If $k = l = i$, 
 
$$
\begin{array}{cl}
(\operatorname{LHS})_{i, i}=& s_i^{\dag}
(\lambda^{i+1, i+1}H (g_{i+1}^{-1}-\lambda)(g_{i+1}-1))
s_i\\
=&\lambda^{i, i}H(g_i^{-1}-\lambda)(g_i-1)\\
=&(\widetilde{H})_{i, i}.
\end{array}
$$
 
If $k = i, l=i+1$,
 
$$
\begin{array}{cl}
&(\operatorname{LHS})_{i, i+1}\\
=& s_i^{\dag}
(\lambda^{i+1, i}H (g_{i+1}^{-1}-1)(g_{i}-1))
s_i g_i\\
&+s_i^{\dag}
(\lambda^{i+1, i+1}H (g_{i+1}^{-1}-\lambda)(g_{i+1}-1))
s_i (1-g_{i+1})\\
=&(\widetilde{H})_{i, i+1}.
\end{array}
$$

If $k = l = i+1$,
$$
\begin{array}{cl}
(\operatorname{LHS})_{i+1, i+1}=& (s_i g_i)^{\dag}\,
(\lambda^{i,i}H (g_{i}^{-1}-\lambda)(g_{i}-1))\,
s_i g_i\\
&+ (s_i g_i)^{\dag}\,
(\lambda^{i,i+1}H (g_{i}^{-1}-1)(g_{i+1}-1))\,
s_i (1-g_{i+1})\\
&+ (s_i(1-g_{i+1}))^{\dag}\,
(\lambda^{i+1,i}H (g_{i+1}^{-1}-1)(g_{i}-1))\,
s_i g_i\\
&+ (s_i(1-g_{i+1}))^{\dag}\,
(\lambda^{i+1,i+1}H (g_{i+1}^{-1}-\lambda)(g_{i+1}-1))\,
s_i (1-g_{i+1}).
\end{array}
$$
Using $s_i^{-1} g_i s_i = g_i g_{i+1} g_i^{-1}$, $s_i^{-1} g_{i+1} s_i = g_i$,
and $g_i^{\dag} H g_i = H$, the first term equals
$\lambda^{-\frac{1}{2}}H(g_{i+1}^{-1}-\lambda)(g_{i+1}-1)
= (\widetilde{H})_{i+1, i+1}$,
while the remaining three terms sum to
$$
\lambda^{-\frac{1}{2}}H(g_{i+1}^{-1}-1)
\left\lbrace (1-g_i^{-1}) - (g_i^{-1}-\lambda)(g_i-1) + \lambda(1-g_i)
\right\rbrace (1-g_{i+1}) = O,
$$
since the bracket vanishes identically. Hence
$(\operatorname{LHS})_{i+1, i+1} = (\widetilde{H})_{i+1, i+1}$.

\end{proof}

\begin{proof}[Proof of Theorem~\ref{thm:non-degenerate}]
 
The proof proceeds in three steps: (1) $K + L \subseteq \mathrm{Ker}\,
\widetilde{H}$; (2) $\mathrm{Ker}\, \widetilde{H} \subseteq K + L$; (3) the
descent of the form and of the unitarity to the quotient.
 
Throughout, we use the following characterization of $L$, due to Haraoka
\cite{haraoka2020multiplicative}, Lemma 5.4:
$\bm{v} = (v_1, \dots, v_n)^{T} \in \mathbb{C}^{Nn}$ belongs to $L$ if and
only if
\begin{equation}\label{eq:char-L}
v_{k} = g_{k+1} v_{k+1} \quad (1 \leq k \leq n-1),
\qquad
v_n = \lambda (g_1 \cdots g_n) v_n.
\end{equation}
 
\emph{Step 1: $K + L \subseteq \mathrm{Ker}\, \widetilde{H}$.}
It suffices to show that $\widetilde{H}K = \bm{0}$ and
$\widetilde{H}L = \bm{0}$.
 
Using $Hg_j^{-1} = g_j^{\dag}H$, which follows from the unitarity of $\rho$
relative to $H$, we have the following equality.
 
$$\widetilde{H} = \lambda^{-\frac{1}{2}}
\begin{pmatrix}
g_1^{\dag} - \lambda& g_1^{\dag} - 1 & \cdots & g_1^{\dag}-1\\
\lambda(g_2^{\dag} - 1)& g_2^{\dag} - \lambda & \cdots & g_2^{\dag} -1\\
\vdots& \vdots & \ddots & \vdots\\
\lambda(g_n^{\dag} - 1)&\lambda (g_n^{\dag} - 1) & \cdots & g_n^{\dag} -\lambda\\    
\end{pmatrix}
H^{\oplus n}
\begin{pmatrix}
g_1 -1 &&\\
&\ddots &\\
&&g_n -1\\    
\end{pmatrix}.$$
 
\noindent By the definition of $K$, $\widetilde{H}K = \bm{0}$.
 
Taking the adjoint of this factorization (recall that $|\lambda| = 1$ and that
$\widetilde{H}^{\dag} = \widetilde{H}$ by Theorem~\ref{thm:uni}), we obtain
 
\begin{equation}\label{eq:second-factorization}
\widetilde{H} = \lambda^{-\frac{1}{2}}
\begin{pmatrix}
g_1^{\dag} -1 &&\\
&\ddots &\\
&&g_n^{\dag} -1\\    
\end{pmatrix}
H^{\oplus n}
\begin{pmatrix}
\lambda g_1 - 1& g_2 - 1 & \cdots & g_n -1\\
\lambda (g_1 - 1)& \lambda g_2 - 1 & \cdots & g_n -1\\
\vdots& \vdots & \ddots & \vdots\\
\lambda (g_1 - 1)& \lambda (g_2 - 1) & \cdots & \lambda g_n -1\\    
\end{pmatrix}.
\end{equation}
 
By the characterization \eqref{eq:char-L} of $L$, every
$\bm{v} = (v_1, \dots, v_n)^{T} \in L$ satisfies $v_{k-1} = g_k v_k$ for
$2 \leq k \leq n$ and $\lambda (g_1 \cdots g_n) v_n = v_n$. Setting
$v_0 := g_1 v_1$, we have $(g_k - 1)v_k = v_{k-1} - v_k$ for all
$1 \leq k \leq n$, and $v_0 = (g_1 \cdots g_n) v_n$. Hence the $j$-th block
row of the right factor of \eqref{eq:second-factorization} applied to $\bm{v}$
telescopes as
\begin{align*}
\lambda \sum_{k=1}^{j-1} (g_k - 1) v_k + (\lambda g_j - 1) v_j
+ \sum_{k=j+1}^{n} (g_k - 1) v_k
&= \lambda (v_0 - v_{j-1}) + (\lambda v_{j-1} - v_j) + (v_j - v_n)\\
&= \lambda (g_1 \cdots g_n) v_n - v_n = 0.
\end{align*}
Therefore $\widetilde{H} L = \bm{0}$, and hence
$K + L \subseteq \mathrm{Ker}\, \widetilde{H}$.
 
\emph{Step 2: $\mathrm{Ker}\, \widetilde{H} \subseteq K + L$.}
Take any $\bm{x} = (x_1, \dots, x_n)^{T} \in \mathbb{C}^{Nn}$, $x_i \in
\mathbb{C}^{N}$, such that $\widetilde{H}\bm{x} = \bm{0}$. From
$g_j^{\dag} H g_j = H$ we obtain
$g_j^{\dag} - 1 = H (g_j^{-1} - 1) H^{-1}$, so that
$\mathrm{Ker}(g_j^{\dag} - 1) = H\, \mathrm{Ker}(g_j - 1)$. Since $H$ is
non-degenerate, the factorization \eqref{eq:second-factorization} therefore
shows that $\widetilde{H}\bm{x} = \bm{0}$ holds if and only if, for each $j$,
the $j$-th block row of the right factor of \eqref{eq:second-factorization}
applied to $\bm{x}$ lies in $\mathrm{Ker}(g_j - 1)$; that is,
\begin{equation}\label{eq:kernel-rows}
\left(\begin{array}{l}
 (\lambda g_1- 1)x_1 + (g_2 -1)x_2 +\cdots + (g_n-1)x_n\\
\phantom{(\lambda g_1- 1)x_1}\vdots\\
 \lambda (g_1- 1)x_1 + \cdots +(\lambda g_j -1)x_j +\cdots + (g_n-1)x_n\\
\phantom{(\lambda g_1- 1)x_1}\vdots\\
 \lambda (g_1- 1)x_1 + \cdots +\lambda (g_{n-1} -1)x_{n-1}  + (\lambda g_n-1)x_n
\end{array}\right)=
\left(\begin{array}{l}
 w_1\\
\vdots\\
 w_j\\
\vdots\\
 w_n
\end{array}\right),
\end{equation}
where $w_j \in \mathrm{Ker}(g_j-1)$.
 
Subtracting the $(j+1)$-th row of \eqref{eq:kernel-rows} from the $j$-th row, we
obtain, for $j = 1, \dots, n-1$,
\begin{equation}\label{eq:kernel-diff}
w_j - w_{j+1} = (\lambda-1)(x_j - g_{j+1}x_{j+1}),
\end{equation}
and hence, since $g_{j+1} w_{j+1} = w_{j+1}$,
$$
(\lambda-1)x_j - w_j = g_{j+1}\left((\lambda-1)x_{j+1} - w_{j+1}\right).
$$

Set $x_j^{'} := (\lambda-1)x_j - w_j$ and
$\bm{x}^{'} := (x^{'}_1, \dots, x^{'}_n)^{T}$. We claim that
$\bm{x}^{'} \in L$; granting this, since
$\bm{w} := (w_1, \dots, w_n)^{T} \in K$, we conclude
$\bm{x} = (\lambda - 1)^{-1}(\bm{x}^{'} + \bm{w}) \in K + L$.
 
The relations above give $x^{'}_j = g_{j+1} x^{'}_{j+1}$ for
$1 \leq j \leq n-1$, so by the characterization \eqref{eq:char-L} it remains
to show $\lambda g_1\cdots g_n x^{'}_n = x^{'}_n$. Multiplying the first row
of \eqref{eq:kernel-rows} by $\lambda$ and subtracting the $n$-th row, all terms
except those in $x_1$ and $x_n$ cancel, and we obtain
$$
\lambda(\lambda - 1)g_1 x_1 - (\lambda - 1)x_n = \lambda w_1 - w_n.
$$
Since $g_1 w_1 = w_1$, this can be rewritten as
$\lambda g_1 x^{'}_1 = x^{'}_n$. Combining this with
$x^{'}_1 = g_2 \cdots g_n x^{'}_n$, we obtain
$\lambda g_1 g_2 \cdots g_n x^{'}_n = x^{'}_n$, as required. Therefore
$\mathrm{Ker}\, \widetilde{H} \subseteq K + L$, and together with Step 1,
$\mathrm{Ker}\, \widetilde{H} = K + L$.
 
\emph{Step 3: descent to the quotient.}
Since $K + L \subseteq \mathrm{Ker}\, \widetilde{H}$ and
$\widetilde{H}^{\dag} = \widetilde{H}$, the value
$\bm{x}^{\dag}\widetilde{H}\bm{y}$ depends only on the classes of $\bm{x}$ and
$\bm{y}$ modulo $K + L$. Hence $\widetilde{H}$ induces a Hermitian form
$\widetilde{H}^{KLM}$ on $V^{\oplus n}/(K+L)$, which is non-degenerate by the
equality $\mathrm{Ker}\, \widetilde{H} = K + L$. Finally, since $K + L$ is
$\rho^{LM}_{\lambda}$-invariant, the invariance
$\rho^{LM}_{\lambda}(\beta)^{\dag}\widetilde{H}\rho^{LM}_{\lambda}(\beta)
= \widetilde{H}$ of Theorem~\ref{thm:uni} descends to
$\rho^{KLM}_{\lambda}(\beta)^{\dag}\widetilde{H}^{KLM}
\rho^{KLM}_{\lambda}(\beta) = \widetilde{H}^{KLM}$ for any
$\beta \in F_n \rtimes B_n$. That is, $\rho^{KLM}_{\lambda}$ is unitary
relative to $\widetilde{H}^{KLM}$.
\end{proof} 

\subsection{Signature of the Hermitian form}\label{Signature}
Subsequently, we determine the signature of $\widetilde{H}$. Here, we introduce the following notation for convenience. For any matrix $X$ and any regular matrix $A$, we define 
$X_A := A^{\dag} X A$; this transformation does not change the signature of $X$. 

\begin{remark}\label{rm:sig}
By Sylvester's law of inertia, the signature of $X_A$ is identical to the signature of $X$.

Moreover, let $U$ be a unitary matrix. Then, the multiset of the eigenvalues of $X_U$ is identical to the multiset of the eigenvalues of $X$.
\end{remark}

In particular, when the matrix defines a Hermitian form, we introduce the following notation. Let $A$ be a $p \times p$ Hermitian matrix and let $X$ be a $p \times q$ matrix; then we denote $X^{\dag}AX$ by $A\lbrack X\rbrack$. Hereafter, whenever we consider unitary matrices, we assume that they have determinant one, i.e., they belong to the special unitary group, $SU(m)$. For simplicity, we will refer to these matrices just as unitary matrices.

\subsubsection{Algorithm to determine the signature of the Hermitian matrix}

Since $\widetilde{H}$ is Hermitian, its eigenvalues are real, and by
Theorem~\ref{thm:non-degenerate} the eigenspace for the eigenvalue $0$, that
is $\mathrm{Ker}\, \widetilde{H}$, coincides with $K + L$. Let $V_{+}$
(resp.\ $V_{-}$) denote the sum of the eigenspaces of $\widetilde{H}$ for the
positive (resp.\ negative) eigenvalues, so that
$$
V^{\oplus n} = V_{+} \oplus V_{-} \oplus (K + L)
$$
is an $\widetilde{H}$-orthogonal decomposition. The quotient map restricts to
an isomorphism $V_{+} \oplus V_{-} \xrightarrow{\ \sim\ } V^{\oplus n}/(K+L)$
which carries the induced form $\widetilde{H}^{KLM}$ to the restriction of
$\widetilde{H}$ to $V_{+} \oplus V_{-}$. Hence the signature of
$\widetilde{H}^{KLM}$ equals $(\widetilde{p}, \widetilde{q})$, where
$\widetilde{p}$ and $\widetilde{q}$ are the numbers of positive and negative
eigenvalues of $\widetilde{H}$ counted with multiplicity. In other words, to
determine the signature of $\widetilde{H}^{KLM}$, it suffices to compute the
signature of $\widetilde{H}$ and discard the zero eigenvalues.

We now construct an algorithm to determine the signature of $\widetilde{H}$. At each step, we view the remaining matrix as a $2\times 2$ block matrix and apply a block-diagonalization procedure recursively. Since the pivot block may be singular, we first establish the following results concerning the kernels and eigenspaces of the component block matrices.

To block-diagonalize $\widetilde{H}$, we introduce the following lemma, which provides a basis adapted to the kernel of $H_1-H_2$, where $H_1$ and $H_2$ are Hermitian matrices.
 
\begin{lemma}\label{lem:kernel-basis}
Let $M=H_1-H_2$ be a $p \times p$ Hermitian matrix. Let $u_{1,0}$ be an orthonormal basis of $\ker M$ and let $u_{1,1}$ be an orthonormal basis of $\operatorname{ran} M = (\ker M)^{\perp}$, and set $U := (u_{1,1}\ \ u_{1,0})$. Then
$$
M \lbrack U \rbrack =
\begin{pmatrix}
M\lbrack u_{1,1}\rbrack & O\\
O & O
\end{pmatrix},
$$
and $M\lbrack u_{1,1}\rbrack$ is invertible.
\end{lemma}
 
\begin{proof}
Since $M$ is Hermitian, $\operatorname{ran} M = (\ker M^{\dag})^{\perp} = (\ker M)^{\perp}$, so that $\mathbb{C}^p = \ker M \oplus \operatorname{ran} M$ is an orthogonal decomposition into $M$-invariant subspaces. This gives the block form. Moreover $M\lbrack u_{1,1}\rbrack$ is the restriction of $M$ to $\operatorname{ran} M$, on which $M$ has trivial kernel; hence it is invertible.
\end{proof}

Lemma~\ref{lem:kernel-basis} block-diagonalizes the $(1,1)$-block of
$\widetilde{H^s}$; it does not, however, control the interaction between
$\ker (\widetilde{H^s})_{11}$ and the remaining block columns. We now show
that, under a condition on $\lambda$ which we call the partial-product invertibility (PI) condition, every kernel vector of
$(\widetilde{H^s})_{11}$ annihilates the entire first block column of
$\widetilde{H^s}$, which justifies the zero blocks appearing in process (b)
below.

\begin{assumption}[partial-product invertibility]\label{ass:PI}
Assume that
\begin{equation}\label{eq:PI}
\mathrm{Ker}\left(\lambda\, g_1 g_2 \cdots g_s - I_N\right) = \{0\}
\qquad \text{for all } s = 1, \dots, n-1.
\tag{PI}
\end{equation}
Equivalently, $\lambda^{-1}$ is not an eigenvalue of any of the partial
products $g_1 \cdots g_s$, $1 \leq s \leq n-1$.
\end{assumption}

Note that \eqref{eq:PI} excludes at most $(n-1)N$ values of $\lambda$, and
that the excluded set is explicitly computable from the eigenvalues of the
partial products. No condition is imposed at $s = n$: the case
$\mathrm{Ker}(\lambda g_1 \cdots g_n - I_N) \neq \{0\}$, that is $L \neq \{0\}$,
is covered by Theorem~\ref{thm:signature-algorithm} below.

The key observation is that the leading principal submatrices of
$\widetilde{H}$ are themselves matrices of the same type.

\begin{lemma}[principal submatrices]\label{lem:principal-submatrix}
For $1 \leq s \leq n$, let $\widetilde{H}^{[s]}$ denote the leading principal
$sN \times sN$ block submatrix of $\widetilde{H}$. Then $\widetilde{H}^{[s]}$
coincides with the matrix $\widetilde{H}$ associated with the truncated tuple
$(g_1, \dots, g_s)$, with the same $H$ and the same $\lambda$.
\end{lemma}

\begin{proof}
The $(j,k)$-block $\lambda^{jk} H (g_j^{-1} - \lambda_{jk} I)(g_k - 1)$
depends only on $g_j$, $g_k$, and on the comparison between $j$ and $k$.
Hence, for $j, k \leq s$, it is unchanged when the tuple is truncated to
$(g_1, \dots, g_s)$.
\end{proof}

The identity $\mathrm{Ker}\, \widetilde{H} = K + L$ of
Theorem~\ref{thm:non-degenerate} is a statement of linear algebra about a
tuple of invertible matrices, the non-degenerate Hermitian matrix $H$, and
$\lambda$: its proof uses only the block formula for $\widetilde{H}$ and the
characterization of $L$ from \cite{haraoka2020multiplicative}, Lemma 5.4.
It therefore applies verbatim to each truncated tuple, and we obtain the
following.

\begin{corollary}\label{cor:subtuple-kernel}
For $1 \leq s \leq n$,
$$
\mathrm{Ker}\, \widetilde{H}^{[s]} = K^{[s]} + L^{[s]},
$$
where $K^{[s]}$ and $L^{[s]}$ denote the subspaces $K$ and $L$ associated
with the tuple $(g_1, \dots, g_s)$; in particular
$\dim L^{[s]} = \dim \mathrm{Ker}(\lambda g_1 \cdots g_s - I_N)$.
Under Assumption~\ref{ass:PI}, for $1 \leq s \leq n-1$ we have
$L^{[s]} = \{0\}$ and hence
\begin{equation}\label{eq:subtuple-count}
\dim \mathrm{Ker}\, \widetilde{H}^{[s]}
= \dim K^{[s]}
= \sum_{k=1}^{s} \dim \mathrm{Ker}(g_k - 1).
\end{equation}
\end{corollary}

The next lemma requires no assumption on $\lambda$: vectors coming from $K$
annihilate entire block columns throughout the recursion. For $1 \leq i \leq s$,
the original $s$-th block occupies the column index $s - i + 1$ of
$\widetilde{H^i}$.

\begin{lemma}[column annihilation]\label{lem:column-annihilation}
Let $1 \leq s \leq n$ and $v \in \mathrm{Ker}(g_s - 1)$. Then, for every
$1 \leq i \leq s$ and every $j$,
$$
(\widetilde{H^i})_{j,\, s-i+1}\, v = 0.
$$
In particular, $\mathrm{Ker}(g_s - 1) \subseteq \mathrm{Ker}\, (\widetilde{H^s})_{11}$.
\end{lemma}

\begin{proof}
We use induction on $i$. For $i = 1$, by the definition of $\widetilde{H}$,
$$
(\widetilde{H})_{js}\, v
= \lambda^{js} H (g_j^{-1} - \lambda_{js} I)(g_s - 1)\, v = 0
\qquad \text{for all } j.
$$
Suppose the claim holds for $i < s$. By the recursion of process (c),
$$
(\widetilde{H^{i+1}})_{j,\, s-i}
= (\widetilde{H^i})_{j+1,\, s-i+1}
- \left((\widetilde{H^i})_{1,\, j+1}\right)^{\dag} P^i\,
  (\widetilde{H^i})_{1,\, s-i+1},
$$
and both terms annihilate $v$ by the induction hypothesis (the second one
because $(\widetilde{H^i})_{1,\, s-i+1} v = 0$).
\end{proof}

We now determine the kernels of the pivots
$(\widetilde{H^s})_{11}$ completely.

\begin{lemma}[pivot kernels]\label{lem:pivot-kernel}
Under Assumption~\ref{ass:PI}:
\begin{enumerate}
\item[(i)] For $1 \leq s \leq n-1$,
$$
\mathrm{Ker}\, (\widetilde{H^s})_{11} = \mathrm{Ker}(g_s - 1),
\qquad \text{so that} \qquad
p_s = N - \dim \mathrm{Ker}(g_s - 1).
$$
\item[(ii)] For $1 \leq s \leq n-1$, every
$u \in \mathrm{Ker}\, (\widetilde{H^s})_{11}$ satisfies
$(\widetilde{H^s})_{j1}\, u = 0$ for all $j$. Consequently, in process (b)
the third block row and column of
$\widetilde{H^s}_{\widetilde{U^s}}$ vanish identically, and the congruence
step of process (c) is valid.
\item[(iii)] At the final step,
$$
\dim \mathrm{Ker}\, (\widetilde{H^n})_{11}
= \dim \mathrm{Ker}(g_n - 1) + \dim \mathrm{Ker}(\lambda g_1 \cdots g_n - I_N).
$$
\end{enumerate}
\end{lemma}

\begin{proof}
For $1 \leq i \leq n-1$, let $W^i$ denote the block unipotent upper triangular
matrix of size $Nn$ which is the identity except that its block row $i$ carries
the entries
$$
(W^i)_{i,\, j} = -\, P^i\, (\widetilde{H^i})_{1,\, j-i+1},
\qquad j = i+1, \dots, n,
$$
where $P^i = u^i_{1,1}\bigl((\widetilde{H^i})_{11}\lbrack u^i_{1,1}\rbrack\bigr)^{-1}
{u^i_{1,1}}^{\dag}$ is as in process (c). A direct computation with the
Hermitian block matrix
$M = \begin{pmatrix} A & B \\ B^{\dag} & C \end{pmatrix}$,
$A = (\widetilde{H^i})_{11}$, shows that the congruence by
$\begin{pmatrix} I & -PB \\ O & I \end{pmatrix}$ yields
\begin{equation}\label{eq:one-step}
\begin{pmatrix}
A & (I - AP)\, B \\
B^{\dag} (I - PA) & C - B^{\dag} P B
\end{pmatrix},
\end{equation}
where $AP$ is the orthogonal projection onto $\operatorname{ran} A$
(Lemma~\ref{lem:kernel-basis}). Hence the off-diagonal remainder
$(I - AP) B$ has rows supported on $\mathrm{Ker}\, A$, and it vanishes if and
only if $u^{\dag} B = 0$ for every $u \in \mathrm{Ker}\, A$; by the Hermitian
symmetry this is the condition that every $u \in \mathrm{Ker}\,(\widetilde{H^i})_{11}$
annihilates the first block column of $\widetilde{H^i}$.

We prove (i) and (ii) simultaneously by induction on $s$.

\emph{Base case $s = 1$.}
Since $H$ is invertible and, by \eqref{eq:PI} with $s = 1$,
$(g_1^{-1} - \lambda I)$ is invertible, we have
$\mathrm{Ker}\, (\widetilde{H})_{11}
= \mathrm{Ker}\, \lambda^{-\frac12} H (g_1^{-1} - \lambda I)(g_1 - 1)
= \mathrm{Ker}(g_1 - 1)$, which gives (i); (ii) follows from
Lemma~\ref{lem:column-annihilation}.

\emph{Inductive step.} Let $2 \leq s \leq n$ and suppose (i), (ii) hold for
all $i < s$. By (ii) for $i < s$, the remainders in \eqref{eq:one-step} vanish
at every step $i < s$, so that $T_{s-1} := W^1 W^2 \cdots W^{s-1}$ satisfies
\begin{equation}\label{eq:flag-congruence}
T_{s-1}^{\dag}\, \widetilde{H}\, T_{s-1}
= (\widetilde{H^1})_{11} \oplus (\widetilde{H^2})_{11} \oplus \cdots \oplus
  (\widetilde{H^{s-1}})_{11} \oplus \widetilde{H^s},
\end{equation}
where the summands occupy the block positions $1, \dots, s-1$ and
$s, \dots, n$ respectively. Each $W^i$ is block unipotent upper triangular,
hence so is $T_{s-1}$; therefore the leading principal $sN \times sN$
submatrix of the left-hand side of \eqref{eq:flag-congruence} equals
$\bigl(T_{s-1}^{[s]}\bigr)^{\dag}\, \widetilde{H}^{[s]}\, T_{s-1}^{[s]}$, where
$T_{s-1}^{[s]}$ is the (invertible) leading principal submatrix of $T_{s-1}$.
Taking the leading principal submatrix of the right-hand side, we obtain the
congruence
\begin{equation}\label{eq:restricted}
\widetilde{H}^{[s]}
\;\cong\;
(\widetilde{H^1})_{11} \oplus \cdots \oplus (\widetilde{H^{s-1}})_{11}
\oplus (\widetilde{H^s})_{11}.
\end{equation}
Comparing kernel dimensions in \eqref{eq:restricted} and using (i) for
$i < s$,
$$
\dim \mathrm{Ker}\, \widetilde{H}^{[s]}
= \sum_{i=1}^{s-1} \dim \mathrm{Ker}(g_i - 1)
+ \dim \mathrm{Ker}\, (\widetilde{H^s})_{11}.
$$
If $s \leq n-1$, Corollary~\ref{cor:subtuple-kernel} gives
$\dim \mathrm{Ker}\, \widetilde{H}^{[s]} = \sum_{i=1}^{s} \dim \mathrm{Ker}(g_i - 1)$,
whence $\dim \mathrm{Ker}\, (\widetilde{H^s})_{11} = \dim \mathrm{Ker}(g_s - 1)$.
Together with the inclusion
$\mathrm{Ker}(g_s - 1) \subseteq \mathrm{Ker}\, (\widetilde{H^s})_{11}$ of
Lemma~\ref{lem:column-annihilation}, this proves (i) at $s$; then (ii) at $s$
follows from Lemma~\ref{lem:column-annihilation} applied to the elements of
$\mathrm{Ker}\,(\widetilde{H^s})_{11} = \mathrm{Ker}(g_s - 1)$.
If $s = n$, Corollary~\ref{cor:subtuple-kernel} (applied without
\eqref{eq:PI}, together with $K \cap L = \{0\}$, see the proof of
Theorem~\ref{thm:signature-algorithm}) gives
$\dim \mathrm{Ker}\, \widetilde{H}^{[n]}
= \sum_{i=1}^{n} \dim \mathrm{Ker}(g_i - 1)
+ \dim \mathrm{Ker}(\lambda g_1 \cdots g_n - I_N)$, which yields (iii).
\end{proof}

Here, we construct the finite sequences of matrices $\widetilde{H^{s}}$, $U^{s}$, for $s=1, \dots , n$ in the following way.

For $s=1$, $\widetilde{H^{1}}=\widetilde{H}$.

For $1\leq s \leq n-1$, we diagonalize $\widetilde{H^s}$ by the following procedure.

\textbf{Diagonalize process (a):}

Calculate the pairs of eigenvalues and the eigenvectors $$(\zeta^s_1, u^s_1), \dots, (\zeta^s_{p_s}, u^s_{p_s}), (0, u^s_{p_s+1}), \dots, (0, u^s_{N})$$ of $(\widetilde{H^s})_{11}$, where $(\widetilde{H^s})_{11}$ is  $N\times N$ Hermitian matrix and $\zeta^s_1, \dots, \zeta^s_{p_s}$ are its nonzero eigenvalues, so that $p_s = \operatorname{rank}(\widetilde{H^s})_{11}$.

\textbf{Diagonalize process (b):}

Define $\widetilde{U^s}$ as $\begin{pmatrix}
    u^s_{1,1}&O&u^s_{1,0}\\
    O&I_{N(n-s)}&O\end{pmatrix}$ where $u^s_{1,1}=(u^s_1\ \dots\ u^s_{p_s})$ is a basis of $\operatorname{ran}(\widetilde{H^s})_{11}$ and $u^s_{1,0}=(u^s_{p_s+1}\ \dots\ u^s_{N})$ is a basis of $\ker(\widetilde{H^s})_{11}$. Then, $\widetilde{H^s}_{\widetilde{U^s}}=
    \begin{pmatrix}
    \widetilde{H^s}_{11}\lbrack u^s_{1,1} \rbrack & {u^s_{1,1}}^{\dag} \widetilde{H^s}_{12}&O\\
    \widetilde{H^s}_{21}u^s_{1,1} & \widetilde{H^s}_{22}&O\\
    O&O&O
    \end{pmatrix}$.
By Lemma~\ref{lem:kernel-basis}, $\widetilde{H^s}_{11}\lbrack u^s_{1,1} \rbrack$ is an invertible $p_s\times p_s$ matrix.

The vanishing of the third block row and column is guaranteed by Lemma~\ref{lem:pivot-kernel} below.

\textbf{Diagonalize process (c):}

 $\left(
    \begin{array}{cc}
    \widetilde{H^s}_{11}\lbrack u^s_{1,1} \rbrack & {u^s_{1,1}}^{\dag} \widetilde{H^s}_{12}\\
    \widetilde{H^s}_{21}u^s_{1,1} & \widetilde{H^s}_{22}\\
    \end{array}
    \right)
    $
    is diagonalized by $$T^s:=\begin{pmatrix}
    I_{p_s}&-(\widetilde{H^s}_{11}\lbrack u^s_{1,1}\rbrack)^{-1} {u^s_{1,1}}^{\dag}(\widetilde{H^s})_{1,2}  \\
    O&I_{N(n-s)} \end{pmatrix}$$ and the result is

$$
\begin{pmatrix}
    H^s_{11}\lbrack u^s_{1,1}\rbrack&O&\cdots&O\\
    O&H^s_{22}- {H^s_{12}}^{\dag}P^s H^s_{12}  &\cdots&H^s_{2,n} - {H^s_{12}}^{\dag}P^s H^s_{1n}\\
    \vdots &\vdots&\ddots&\vdots\\
    O &H^s_{n,2}- {H^s_{1n}}^{\dag}P^s H^s_{12}&\cdots&H^s_{n, n} - {H^s_{1n}}^{\dag}P^s H^s_{1n}
\end{pmatrix},$$ 
where $H^s_{kl}$ is the $(k,l)$-th block of $\widetilde{H^s}$, and where we abbreviate
$$
P^s := u^s_{1,1}\left(H^s_{11}\lbrack u^s_{1,1}\rbrack\right)^{-1}{u^s_{1,1}}^{\dag}.
$$
Note that $H^s_{11}$ itself may be singular, so that $(H^s_{11})^{-1}$ need not exist; only its restriction $H^s_{11}\lbrack u^s_{1,1}\rbrack$ to $\operatorname{ran}H^s_{11}$ is inverted, and $u^s_{1,1}$ transports the result back. Since $\widetilde{H^s}$ is a Hermitian matrix, $H_{kl}^{\dag} = H_{lk}$, and $P^s$ is Hermitian; hence each $\widetilde{H^{s+1}}$ below is again Hermitian.

Then we define the $(n-s)\times (n-s)$ block matrix $\widetilde{H^{s+1}}$ as

$$
\begin{pmatrix}
    H_{22}- H_{12}^{\dag}P^s H_{12}  &\cdots&H_{2n} - H_{12}^{\dag}P^s H_{1n}\\
    \vdots&\ddots&\vdots\\
    H_{n,2}- H_{1n}^{\dag}P^s H_{12}&\cdots&H_{nn} - H_{1n}^{\dag}P^s H_{1n}
\end{pmatrix}.
$$

Combining the discussion above, we obtain the following.

\begin{theorem}\label{thm:signature-algorithm}
Suppose Assumption~\ref{ass:PI} holds. Then
$W := W^1 W^2 \cdots W^{n-1}$ is invertible and
\begin{equation}\label{eq:full-congruence}
W^{\dag}\, \widetilde{H}\, W
= (\widetilde{H^1})_{11} \oplus (\widetilde{H^2})_{11} \oplus \cdots \oplus
  (\widetilde{H^n})_{11},
\end{equation}
where the blocks $(\widetilde{H^s})_{11}$ are produced by the diagonalization
processes (a)--(c). Consequently:
\begin{enumerate}
\item[(i)] The signature $(\widetilde{p}, \widetilde{q})$ of $\widetilde{H}$
is given by
$\widetilde{p} = \sum_{s=1}^{n} \widetilde{p_s}$,
$\widetilde{q} = \sum_{s=1}^{n} \widetilde{q_s}$,
where $(\widetilde{p_s}, \widetilde{q_s})$ is the signature of the invertible
matrix $(\widetilde{H^s})_{11}\lbrack u^s_{1,1} \rbrack$.
\item[(ii)] $\operatorname{rank} \widetilde{H} = \sum_{s=1}^{n} p_s$, with
$p_s = N - \dim\mathrm{Ker}(g_s - 1)$ for $s \leq n-1$ and
$p_n = N - \dim\mathrm{Ker}(g_n - 1) - \dim\mathrm{Ker}(\lambda g_1 \cdots g_n - I_N)$.
\item[(iii)] The multiplicity of the eigenvalue $0$ of $\widetilde{H}$ equals
$$
\dim (K + L)
= \sum_{s=1}^{n} \dim \mathrm{Ker}(g_s - 1)
+ \dim \mathrm{Ker}(\lambda g_1 \cdots g_n - I_N),
$$
consistently with Theorem~\ref{thm:non-degenerate}; in particular, the Hermitian
form induced by $\widetilde{H}$ on $V^{\oplus n} / (K + L)$ is non-degenerate with
signature $(\widetilde{p}, \widetilde{q})$.
\end{enumerate}
\end{theorem}

\begin{proof}
By Lemma~\ref{lem:pivot-kernel} (ii), the remainders in \eqref{eq:one-step}
vanish at every step $s \leq n-1$, which gives \eqref{eq:full-congruence};
note that the last step requires no condition, since there is no block column
below the $n$-th. Statement (i) follows from Sylvester's law of inertia
together with Lemma~\ref{lem:kernel-basis}, and (ii) from
Lemma~\ref{lem:pivot-kernel} (i), (iii). For (iii), it remains to note that
$K \cap L = \{0\}$ when $\lambda \neq 1$: if
$\bm{x} = (w_1, \dots, w_n)^{T} \in K \cap L$, then $w_i \in \mathrm{Ker}(g_i - 1)$
and, by the characterization \eqref{eq:char-L} of $L$,
$w_i = g_{i+1} w_{i+1} = w_{i+1}$ for all $i$; hence all the $w_i$ are equal
to a common vector $w$, which satisfies $w = \lambda (g_1 \cdots g_n) w = \lambda w$,
so $w = 0$. Therefore
$\dim(K + L) = \dim K + \dim L$, and the count of zero eigenvalues in
\eqref{eq:full-congruence} matches $\dim(K+L)$ by
Theorem~\ref{thm:non-degenerate}.
\end{proof}

\begin{proposition}[extension beyond the PI condition]\label{prop:PI-removal}
Let $|\lambda_0| = 1$, $\lambda_0 \neq 1$, and suppose that $\lambda_0^{-1}$
is not an eigenvalue of the full product $g_1 g_2 \cdots g_n$. Fix a
continuous branch of $\lambda^{\frac12}$ on a neighborhood of $\lambda_0$ in
the unit circle containing the chosen square root. Then the signature
$(\widetilde{p}, \widetilde{q})$ of $\widetilde{H}(\lambda)$ is constant on a
neighborhood $U$ of $\lambda_0$ in the unit circle. Since all but finitely
many $\lambda \in U$ satisfy \eqref{eq:PI}, the signature at $\lambda_0$ is
computed by applying Theorem~\ref{thm:signature-algorithm} at any such
$\lambda$. Thus, even when $\lambda_0$ does not satisfy
Assumption~\ref{ass:PI}, the signature of $\widetilde{H}(\lambda_0)$ is
obtained by applying Theorem~\ref{thm:signature-algorithm} at any sufficiently
close parameter $\lambda\in U$ satisfying Assumption~\ref{ass:PI}.
\end{proposition}

\begin{proof}
With the chosen branch of $\lambda^{\frac12}$, the entries of
$\widetilde{H}(\lambda)$ depend continuously on $\lambda$. By
Theorem~\ref{thm:non-degenerate}, which does not require
Assumption~\ref{ass:PI}, we have
$\mathrm{Ker}\, \widetilde{H}(\lambda) = K + L(\lambda)$, where $\dim K$ is
independent of $\lambda$ and
$\dim L(\lambda) = \dim \mathrm{Ker}(\lambda\, g_1 \cdots g_n - I_N)$. Since
the spectrum of $g_1 \cdots g_n$ is finite and
$\lambda_0^{-1} \notin \mathrm{spec}(g_1 \cdots g_n)$, there is a
neighborhood $U$ of $\lambda_0$ in the unit circle on which
$L(\lambda) = \{0\}$, so that
$\dim \mathrm{Ker}\, \widetilde{H}(\lambda) = \dim K$ is constant on $U$.
The eigenvalues of the Hermitian matrix $\widetilde{H}(\lambda)$ vary
continuously in $\lambda$; if the number of positive eigenvalues were not
locally constant on $U$, some eigenvalue would vanish at a point of $U$,
increasing the multiplicity of the eigenvalue $0$ beyond $\dim K$, a
contradiction. Hence $(\widetilde{p}, \widetilde{q})$ is constant on $U$.
Finally, Assumption~\ref{ass:PI} fails only on the finite set
$\bigcup_{s=1}^{n-1} \{\lambda;\ \lambda^{-1} \in
\mathrm{spec}(g_1 \cdots g_s)\}$, so $U$ contains points satisfying
\eqref{eq:PI} arbitrarily close to $\lambda_0$, and the claim follows.
\end{proof}

\begin{remark}\label{rm:arcs}
By Proposition~\ref{prop:PI-removal}, the signature
$(\widetilde{p}, \widetilde{q})$ is locally constant, hence constant, on each
connected component of
$\{\lambda;\ |\lambda| = 1\} \setminus (\{1\} \cup E)$, where
$E = \{\lambda;\ |\lambda| = 1,\
\det(\lambda\, g_1 \cdots g_n - I_N) = 0\}$; it may vary from one
component to another. This observation is taken up again in the Discussion.
\end{remark}

\begin{remark}[role of the PI condition in the blockwise recursion]\label{rm:PI-role}
If Assumption~\ref{ass:PI} fails, the pivot $(\widetilde{H^s})_{11}$ may
acquire additional kernel directions arising from $L^{[s]}$. The argument
proving the vanishing of the off-diagonal remainder then no longer applies,
so the direct blockwise recursion is not guaranteed to work at that
parameter. Nevertheless, Proposition~\ref{prop:PI-removal} shows that, away
from the exceptional set associated with the full product $g_1\cdots g_n$,
the signature can still be computed at a nearby parameter satisfying
Assumption~\ref{ass:PI}. The excluded values may genuinely occur when
$\lambda$ and the eigenvalues of the $g_k$ are roots of unity.
\end{remark}

\begin{remark}[dependence on the ordering]\label{rm:ordering}
The excluded set in Assumption~\ref{ass:PI} depends on the ordering
of the tuple $(g_1, \dots, g_n)$ through the partial products, whereas the
signature of $\widetilde{H}$ does not. Tuples related by the braid group
action yield equivalent representations, so one expects that
Assumption~\ref{ass:PI} may be replaced by the requirement that
\eqref{eq:PI} holds for some tuple in the Hurwitz orbit of
$(g_1, \dots, g_n)$; we do not pursue this here.
\end{remark}

\section{Discussion}

In this paper, we compared the Katz--Long--Moody construction with
Haraoka's multiplicative middle convolution for KZ-type equations.
After identifying $P_{n+1}$ with $F_n\rtimes P_n$ and passing between
representations and anti-representations, the Haraoka--Long correspondence
of Theorem~\ref{haraoka-long} identifies the restriction of the twisted
Long--Moody construction to $P_{n+1}$ with Haraoka's convolution.
This correspondence provides an algebraic interpretation of the
multiplicative middle convolution and allows properties of the two
constructions to be compared directly. In particular, under the
irreducibility hypotheses stated above, irreducibility is preserved by
the construction.

The main unitarity result gives a more precise description of the
Hermitian form preserved by the resulting representation. If the initial
representation $\rho$ is unitary relative to a non-degenerate Hermitian
form $H$, then the matrix $\widetilde{H}$ satisfies the corresponding
invariance relation, and Theorem~\ref{thm:non-degenerate} gives
\[
\operatorname{Ker}\widetilde{H}=K+L.
\]
Consequently, $\widetilde{H}$ induces a non-degenerate invariant Hermitian
form $\widetilde{H}^{KLM}$ on
\[
V^{\oplus n}/(K+L).
\]
Thus, the Katz--Long--Moody construction preserves unitarity in the sense
used throughout this paper, namely, unitarity with respect to a
non-degenerate Hermitian form that may be indefinite.

Moreover, under Assumption~\ref{ass:PI},
Theorem~\ref{thm:signature-algorithm} determines the signature of
$\widetilde{H}^{KLM}$ by a finite sequence of block congruences.
The result therefore distinguishes whether the invariant form produced
by the construction is positive definite, negative definite, or
indefinite. From the analytic point of view, this form is an invariant
Hermitian form for the monodromy representation obtained by the
multiplicative middle convolution. Its signature consequently gives an
invariant connecting the algebraic Katz--Long--Moody construction with
the analytic monodromy of KZ-type equations.

We conclude by formulating a related question in our setting.
Birman and Brendle \cite{birman2005braids} record the broad question of
whether finite-dimensional unitary matrix representations of braid groups
arise in a manner related to Long's construction. As in
\cite{long1994constructing}, and in accordance with the terminology of
this paper, unitarity here is understood with respect to a non-degenerate,
possibly indefinite, Hermitian form. Under the hypotheses of
Theorem~\ref{thm:non-degenerate}, the Katz--Long--Moody construction
preserves unitarity in precisely this sense.

A natural refinement is to ask when the invariant form produced by the
construction is definite. Whenever $\widetilde{H}^{KLM}$ is definite,
multiplying it by $-1$ if necessary yields a positive-definite invariant
Hermitian form, and hence the resulting representation is unitary in the
classical positive-definite sense.

\begin{problem}\label{prob:definite}
Let $\rho$ be unitary relative to a positive-definite Hermitian form $H$,
and set $g_i=\rho(x_i)$ for $1\leq i\leq n$.
Determine all parameters $\lambda\in\mathbb{C}$ satisfying
$|\lambda|=1$ and $\lambda\neq1$ for which the induced Hermitian form
$\widetilde{H}^{KLM}$ on
\[
V^{\oplus n}/(K+L)
\]
is definite. More generally, characterize the tuples
$(g_1,\dots,g_n)$ for which such a parameter exists.
\end{problem}

For a fixed tuple $(g_1,\dots,g_n)$, the results of this paper reduce
Problem~\ref{prob:definite} to finitely many computations. Indeed, the
subspace $K$ is independent of $\lambda$. Since
\[
\operatorname{Ker}\widetilde{H}=K+L
\qquad\text{and}\qquad
K\cap L=\{0\}
\]
for $\lambda\neq1$, the form induced by $\widetilde{H}$ on
$V^{\oplus n}/K$ is degenerate precisely when $L\neq\{0\}$.
Furthermore,
\[
\dim L
=
\dim\operatorname{Ker}
\bigl(\lambda g_1\cdots g_n-I_N\bigr).
\]
Thus degeneracy occurs only at the finite set
\[
E=
\left\{
\lambda\in\mathbb{C};
\ |\lambda|=1,\ \lambda\neq1,\ 
\det\bigl(\lambda g_1\cdots g_n-I_N\bigr)=0
\right\}.
\]

On each connected component of the unit circle with
$\{1\}\cup E$ removed, the subspace $L$ vanishes. After choosing a
continuous branch of $\lambda^{1/2}$ on that component, the entries of
the induced non-degenerate Hermitian form vary continuously with
$\lambda$. Its signature is therefore constant on the component
(Proposition~\ref{prop:PI-removal}).
Replacing $\lambda^{1/2}$ by $-\lambda^{1/2}$ replaces
$\widetilde{H}$ by $-\widetilde{H}$ and interchanges the positive and
negative indices; hence definiteness is independent of the choice of
square root.

Assumption~\ref{ass:PI} excludes only finitely many additional
values of $\lambda$, arising from the partial products
$g_1\cdots g_s$ with $s\leq n-1$. Consequently, one may choose a
sample point satisfying the PI condition \eqref{eq:PI} in each of the open arcs determined by
$\{1\}\cup E$ and compute its signature using the algorithm of
Section~\ref{Signature}. At each of the finitely many parameters in $E$, the signature can be
computed directly by diagonalizing $\widetilde{H}(\lambda)$ and discarding
its kernel $K+L$; at such points the quotient by $K+L$ may have smaller
dimension. In this way, Problem~\ref{prob:definite} can be decided for any
fixed tuple by finitely many computations.

The excluded values arising from the proper partial products do not
by themselves change the signature of the full non-degenerate form,
provided that $\lambda\notin E$; rather, they obstruct the direct use of
the recursive block-diagonalization algorithm at those particular
parameters. Extending the algorithm to such parameters and
deriving a formula for the change of signature at the points of $E$
remain natural problems for further study. Results on simultaneous
eigenspace decompositions for middle convolution and monodromy, such as
those of Oshima \cite{Oshima2016TRANSFORMATIONSOK}, may be useful in this
direction.

In the rank-one case $N=1$, the representations and invariant forms are
closely related to those of Burau--Gassner type and to classical
hypergeometric monodromy \cite{delignemonodromy}. This case should provide
a useful testing ground for explicit criteria for definiteness and for
a wall-crossing formula describing how the signature changes as
$\lambda$ passes through an exceptional value.

For rank two Fuchsian systems, a complete characterization of unitary
monodromies, together with the classification of their signatures into the
definite and indefinite cases, has been obtained by
Adachi~\cite{adachi2024unitary}.

\bmhead{Acknowledgements}

The author would like to express her sincere gratitude to Shunya Adachi, Michael Dettweiler, Yoshishige Haraoka, Kazuki Hiroe, the late Kenji Iohara, Christian Kassel, Taro Kimura, Fuyuta Komura, Claude Mitschi, Yasunori Okada, Toshio Oshima, Stefan Reiter, and Kouichi Takemura for valuable discussions and insightful comments. The author is especially grateful to the late Professor Kenji Iohara for his generous guidance and encouragement during her visits to Lyon. The author also thanks Kenichi Bannai for his continuous encouragement. This work was supported by JST SPRING, Grant Number JPMJSP2109.

\section*{Declarations}

\bmhead{Funding}
This work was supported by JST SPRING, Grant Number JPMJSP2109.

\bmhead{Competing interests}
The author declares no competing interests.

\bmhead{Data availability}
Data sharing is not applicable to this article as no datasets were generated
or analysed during the current study.

\bibliography{sn-bibliography}

\end{document}